\nonstopmode
\documentclass[12pt]{article}
\usepackage[utf8]{inputenc}
\usepackage{import}
\usepackage[authoryear]{natbib}

\usepackage[hidelinks]{hyperref}
\usepackage{fancyhdr}
\usepackage{amsfonts, amsmath, amsthm, amssymb}
\usepackage{bm}
\usepackage{setspace}
\usepackage{graphicx}
\usepackage[a4paper, margin=1in]{geometry}
\usepackage{appendix}
\usepackage[ruled,vlined]{algorithm2e}
\usepackage{array}
\usepackage{pdfpages}
\usepackage{caption}
\usepackage{subcaption}
\usepackage{listings}
\usepackage{xcolor}
\usepackage{booktabs}
\usepackage{multirow}
\usepackage{tipa}

\usepackage{titlesec} 
\titleformat{\section}{\Large \bfseries}{\thesection}{1em}{}
\titleformat{\subsection}{\large\itshape}{\thesubsection}{1em}{}
\titleformat{\subsubsection}{\itshape}{\thesubsubsection}{1em}{}

\usepackage{soul}

\newcommand{\logit}{\mathrm{logit}}
\newcommand{\ilogit}{\mathrm{ilogit}}

\usepackage{soul}
\makeatletter
\newif\if@blind
\@blindfalse  
\if@blind \def\hl[#1]#2{#1}\else
    \def\hl[#1]#2{#2}\relax
\fi

\title{Exploring British Accents: Modelling the Trap--Bath Split with Functional Data Analysis}
\author{\hl[]{Aranya Koshy and Shahin Tavakoli\footnote{Aranya Koshy was an MMORSE student at the University of Warwick, and Shahin Tavakoli is Assistant Professor of Statistics, University of Warwick, CV4 7AL Coventry. Emails: aranya.koshy@gmail.com, s.tavakoli@warwick.ac.uk}}}
\date{June 14, 2021}

\begin{document} 
\maketitle
\thispagestyle{empty}

\setcounter{page}{1}
\setstretch{1.45}
\renewcommand{\arraystretch}{1.5}

\Large
 \begin{center}
\bf Abstract\\ 
\end{center}
\hspace{10pt}
\normalsize
The sound of our speech is influenced by the places we come from. Great Britain contains a wide variety of distinctive accents which are of interest to linguistics. In particular, the ``a'' vowel in words like ``class'' is pronounced differently in the North and the South. Speech recordings of this vowel can be represented as formant curves or as mel-frequency cepstral coefficient curves. Functional data analysis and generalised additive models offer techniques to model the variation in these curves. Our first aim is to model the difference between typical Northern and Southern vowels /\ae/ and /\textipa{A}/, by training two classifiers on the North-South Class Vowels dataset\hl[]{ collected for this paper}
. Our second aim is to visualise geographical variation of accents in Great Britain. For this we use speech recordings from a second dataset, the British National Corpus (BNC) audio edition
. The trained models are used to predict the accent of speakers in the BNC, and then we model the geographical patterns in these predictions using a soap film smoother. This work demonstrates a flexible and interpretable approach to modelling phonetic accent variation in speech recordings.

\paragraph{Keywords:} Formants, functional principal component analysis, logistic regression,  phonetics,  MFCC.
\newpage

\setstretch{1.5}
\renewcommand{\arraystretch}{1.5}

\section{Phonetics} \label{sec:intro}

Phonetic variation in speech is a complex and fascinating phenomenon. The sound of our speech is influenced by the communities and groups we belong to, places we come from, the immediate social context of speech, and many physiological factors. There is acoustic variation in speech due to sex and gender specific differences in articulation \citep{Huber1999}, age \citep{Safavi2018}, social class and ethnicity \citep{Clayards}, and individual idiosyncrasies of sound production \citep{Noiray2014VariabilityAcoustics}. This linguistic variation is relevant to many fields of study like anthropology, economics and demography
\citep{Ginsburgh2014}, and has connections to the study of speech production and perception in the human brain. It helps us understand how languages developed in the past, and the evolutionary links that still exist between languages today \citep{Pigoli}. Modelling phonetic variation is also important for many practical applications, like speech recognition and speech synthesis. In this work, we study one source of variation in particular: geographical accent variation.

To describe phonetic variation conveniently, in his seminal work \textit{Accents of English}, \citet{wells_1982} introduced lexical sets, which are groups of words containing vowels that are pronounced the same way within an accent. The \textit{trap} lexical set contains words like trap, cat and man, and the \textit{bath} lexical set contains words like bath, class and grass. In Northern English accents both \textit{trap} and \textit{bath} words use the `short a' vowel /\ae/. In Southern English accents \textit{trap} words use /\ae/ and \textit{bath} words use the  `long a' vowel /\textipa{A}/; this is known as the trap-bath split. 

\begin{figure}[h]
    \centering
    \includegraphics[width=2.5in]{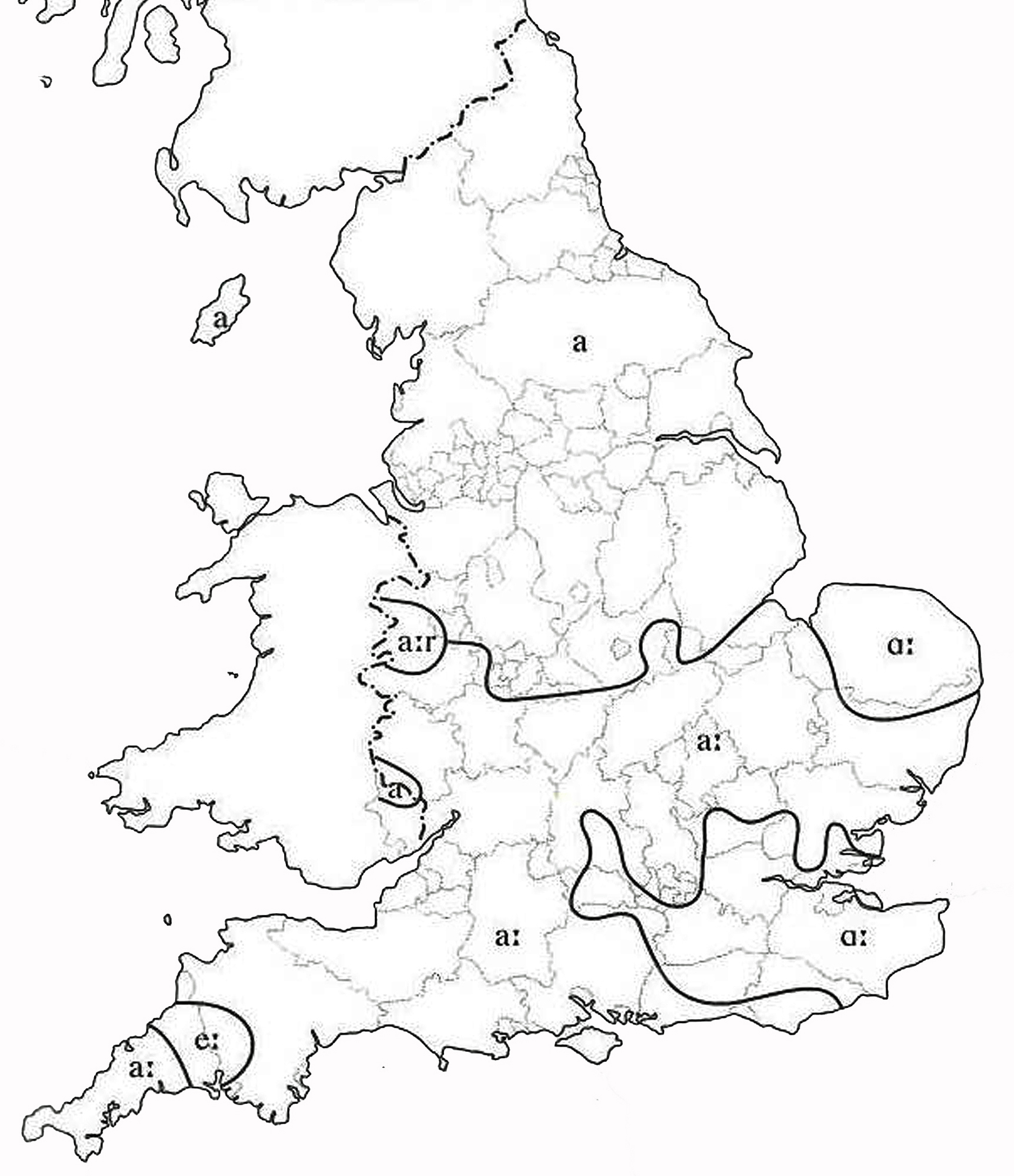}
    \caption{Isoglosses for the ``class'' vowel in England. Reproduced with permission from \citet[][p.\ 6--7]{Upton1996}.}
    \label{fig:isogloss}
\end{figure}

The trap-bath split is one of the most well studied geographical accent differences. The geographical accent variation in sounds like these has historically been studied using written transcriptions of speech from surveys and interviews by trained linguists. These were used to construct isogloss maps (see Figure~\ref{fig:isogloss}) to visualise regions having the same dialect. \citet{Upton1996} explain that in reality these isoglosses are not sharp boundaries, and they are drawn to show only the most prominent linguistic variation in a region for the sake of simplicity. The boundaries are also constantly moving and changing over time.

More recently, advances in statistical methods and technology have allowed accent variation to be modelled by directly using audio recordings of speech. A sound can be represented as a set of smooth curves, and functional data analysis \citep[FDA;][]{Ramsay2005,ferraty:vieu:2006,horvath:2012:book} offers techniques to model variation in these curves. This work demonstrates one such approach, in which we analyse variation in vowel sounds using techniques from FDA and generalised linear models.

This paper has two main contributions. The first contribution is to use functional data analysis to classify vowels by directly using speech recordings: we demonstrate two approaches for classifying \textit{bath} vowels as Northern or Southern. The first approach models variation in formant curves (see Section~\ref{sec:formants}) using a functional linear model. The second approach models variation in mel-frequency cepstral coefficient (MFCC) curves (see Section~\ref{sec:mfcc}) through penalised logistic regression on functional principal components, and it can be used to resynthesise vowel sounds in different accents, allowing us to ``listen to the model''. Both approaches classify accents using the temporal dynamics of the MFCC or formant curves in sounds. These two classifiers were trained using a dataset of labelled audio recordings\hl[]{ that was collected for this paper in an experimental setup} \citep{Koshy2020_shahin}.
The second contribution is to construct maps that visualise geographic variation in the \textit{bath} vowel that can be attributed to typical Northern and Southern accent differences, using a soap film smoother. For this we use the audio BNC dataset \citep{BNC}, which is a representative sample of accents in Great Britain. The resulting maps show a geographical variation in the vowel similar to what is seen in isogloss maps like Figure~\ref{fig:isogloss}.

The paper is structured as follows. In Section~\ref{sec:preprocessing}, we introduce two ways of representing vowel sounds as multivariate curves. Section~\ref{sec:data} introduces the two datasets used in this analysis, and the preprocessing steps involved. Section~\ref{sec:classify} gives the two models for classifying \textit{bath} vowels, and Section~\ref{sec:maps} presents the maps constructed to visualise geographical accent variation. We conclude with a discussion of the results in Section~\ref{sec:discussion}.

\section{Sound as data objects} \label{sec:preprocessing}

\begin{figure}[hb]
    \centering
    \includegraphics[width=3in]{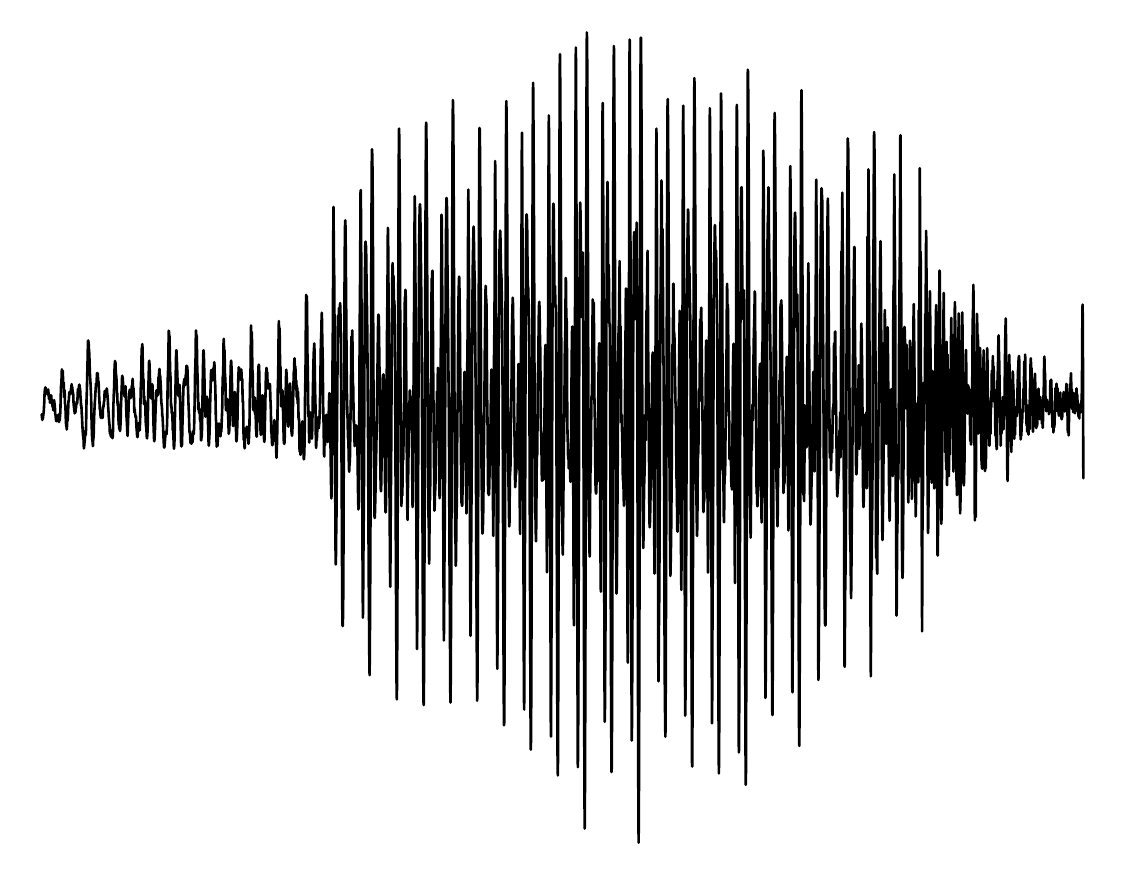}
    \caption{Sound wave of the vowel from a single ``last'' utterance.}
    \label{fig:waveform}
\end{figure}

Sound is a longitudinal air pressure wave. Microphones measure the air pressure at fixed rates, for example at 16 kHz (Hz is a unit of frequency representing samples per second). The waveform of the vowel in the word ``class'' in Figure~\ref{fig:waveform} shows this rapidly oscillating air pressure wave as measured by a microphone. This signal can be transformed in several ways to study it; for example as a spectrogram, formants, or mel-frequency cepstral coefficients (MFCCs), see Sections~\ref{sec:spec}, \ref{sec:formants} and \ref{sec:mfcc}. 

\subsection{Spectrograms}
\label{sec:spec}

\begin{figure}[p]
    \centering
    \includegraphics[width=4in]{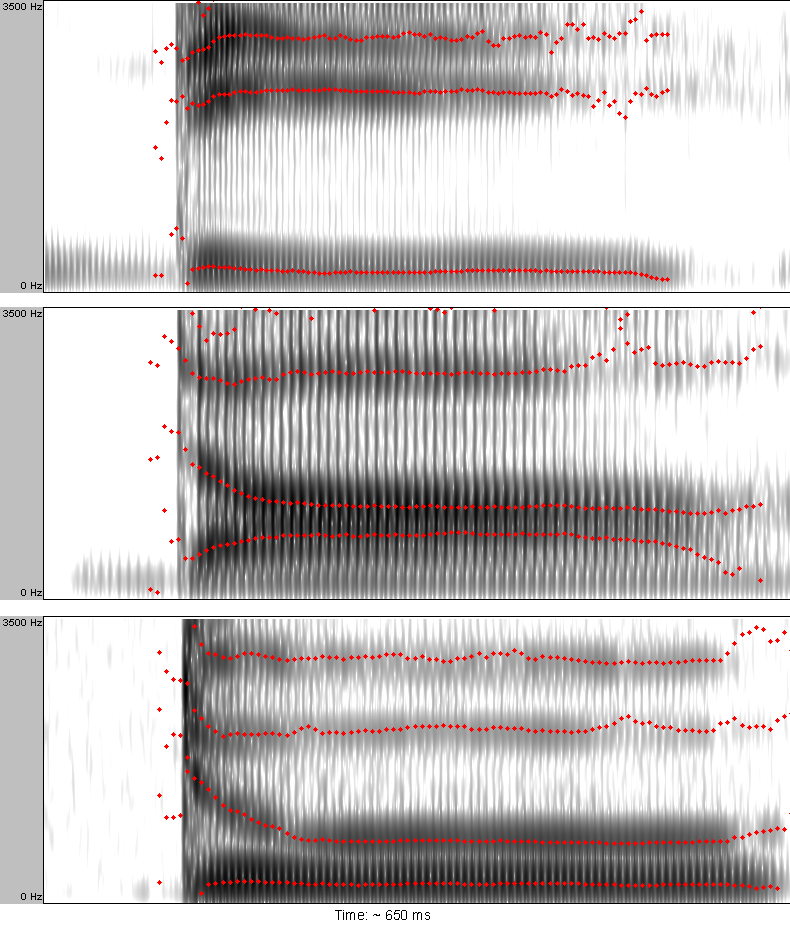}
    \caption{In these spectrograms of the syllables \textit{dee, dah, doo}, the dark
    bands are the formants of each vowel and the overlaid red dotted lines are estimated formant trajectories. The y axis represents frequency and
darkness represents intensity \citep{kluk2007}. }
    \label{fig:formants}
\end{figure}

We begin by defining the spectrogram of a sound. A spectrogram is a time-frequency representation of a sound: it reveals how the most prominent frequencies in a sound change over time. To define it precisely, let us denote the sound wave as a time series $\{s(t): t = 1, \ldots, T\}$, where $s(t)$ is the deviation from normal air pressure at time $t$. We can define $s(t)=0$ for $t\le 0$ or $t>T$. Let $w: \mathbb{R} \rightarrow \mathbb{R}$ be a symmetric window function which is non-zero only in the interval $[-\frac{M}{2},\frac{M}{2}]$ for some $M<T$. The Short-Time Fourier Transform of $\{s(t)\}_{t=1}^T$ is computed as
\begin{align*}
    \text{STFT}(s)(t, \omega) &= \sum_{u=-\infty}^\infty s(u)w(u-t)\text{exp}(-i\omega u) \\ 
    & =  \sum_{u=1}^T s(u)w(u-t)\text{exp}(-i\omega u), 
\end{align*}
for $t=1,\ldots,T$, and $\omega \in \{2\pi k/N: k=0, \ldots, N-1\}$ for some $N\ge T$ which is a power of 2. The window width $M$ is often chosen to correspond to a 20 ms interval. 
The spectrogram of $\{s(t)\}_{t=1}^T$ is then defined as
\begin{align*}
    \text{Spec}(s)(t, \omega) & = |\text{STFT}(s)(t, \omega)|^2.
\end{align*}
At a time point $t$, the spectrogram shows the magnitude of different frequency components $\omega$ in the sound. Figure~\ref{fig:formants} shows spectrograms of recordings of different vowels, with time on the x-axis, frequency on the y-axis, and colour representing the amplitude of each frequency. The dark bands are frequency peaks in the sound, which leads us to the concept of formants.

\subsection{Formants}
\label{sec:formants}
Formants are the strongest frequencies in a vowel sound, observed as high-intensity bands in the spectrogram of the sound. By convention they are numbered in order of increasing frequency, $\text{F}_1, \text{F}_2, \ldots$. 

\begin{figure}[h]
    \centering
    \includegraphics[width=2in]{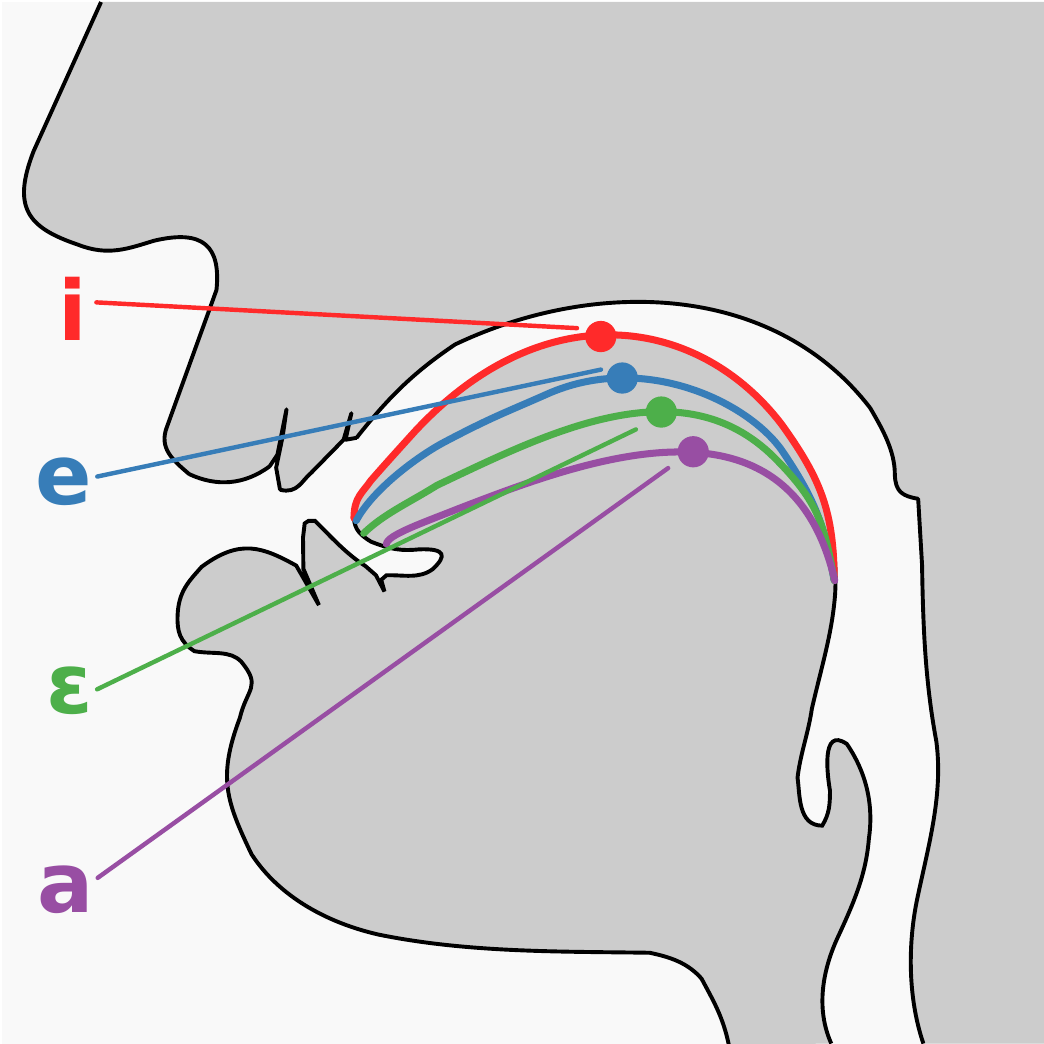}
    \caption{This diagram shows how varying the height of the tongue creates different vowels \citep{CC2008_shahin}.}
    \label{fig:vocaltract}
\end{figure}

Formants are produced by the resonating cavities and tissues of the vocal tract \citep{Johnson2005}. The resonant frequencies depend on the shape of the vocal tract, which is influenced by factors like rounding of the lips, and height and shape of the tongue (illustrated in Figure~\ref{fig:vocaltract}). The pattern of these frequencies is what distinguishes different vowels. They are particularly important for speech perception because of their connection to the vocal tract itself, and not the vocal cords. Listeners use formants to identify vowels even when they are spoken at different pitches, or when the vowels are whispered and the vocal cords don't vibrate at all \citep{Johnson2005}. One can also sometimes ``hear'' a person smile as they speak, because the act of smiling changes the shapes of the vocal cavities and hence the formants produced \citep{Ponsot2018}.

\subsection{Mel-Frequency Cepstral Coefficients}
\label{sec:mfcc}

Mel-frequency cepstral coefficients (MFCCs) are a further transformation of the spectrogram, and are often used in speech recognition and speech synthesis. The way they are constructed is related to how the human auditory system processes acoustic input; in particular, how different frequency ranges are filtered through the cochlea in the inner ear. This filtering is the reason humans can distinguish between low frequencies better than high frequencies. MFCCs roughly correspond to the energy contained in different frequency bands, but are not otherwise easily interpretable.
There are many variants of MFCCs; we use the one from \citet{Erro2011,Erro2014} which allow for high fidelity sound resynthesis.

\begin{figure}[h]
    \centering
    \includegraphics[height=3in]{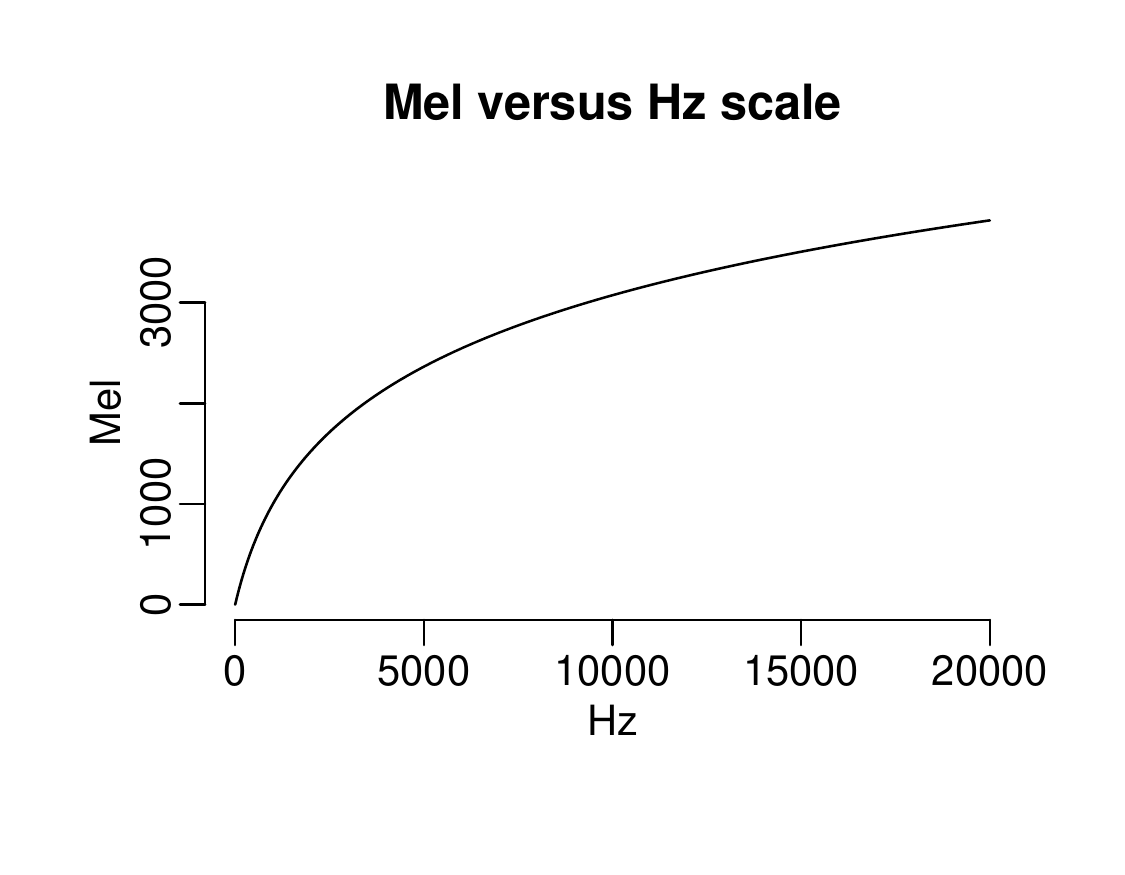}
    \caption{Mapping from Hz to mel. A pair of high frequencies on the Hz scale sound more similar to the human ear than an equidistant pair at low frequencies. This is captured by the mel scale.}
    \label{fig:melfilter}
\end{figure}

MFCCs are computed in two steps as follows \citep{Tavakoli2019}. First the mel-spectrogram is computed from the spectrogram, using a mel scale filter bank with $F$ filters $(b_{f,k})_{k=0,\ldots,N-1}$, $f=0, \ldots, F$. The mel scale is a perceptual scale of pitches, under which pairs of sounds that are perceptually equidistant in pitch are also equidistant in mel units. 
This is unlike the linear Hz scale, in which a pair of low frequencies will sound further apart than an equidistant pair of high frequencies. 
The mapping from Hz ($f$) to mels ($m$) is given by $m=2595\, \text{log}_{10}(1+f/700)$, shown in Figure~\ref{fig:melfilter}. The mel-spectrogram is defined as
\begin{align*}
    \text{MelSpec}(s)(t, f) &= \sum_{k=0}^{N-1} \text{Spec}(s)(t, 2\pi k/N)b_{f,k}.
\end{align*}
In the second step, we take the inverse Fourier transform of the logarithm of this mel-spectrogram. The first $M$ resulting coefficients are the MFCCs,
\begin{align*}
    \text{MFCC}(s)(t, m) &= \frac{1}{F}\sum_{f=0}^{F} \text{log}\left(\text{MelSpec}(s)(t, f)\right) \text{exp} \left( i\frac{2\pi (m-1)f}{F+1} \right).
\end{align*}
At each time point $t$ we have $M$ MFCCs. We use the \texttt{ahocoder} software \citep{Erro2014} to extract MFCCs, which uses $M=40$ at each time point. Thus we represent each vowel sound by 40 MFCC curves.

Formants are a low-dimensional summary of the original sound which allow interpretation of the vocal tract position. MFCCs retain a lot of information about speech sounds and do not simplify the representation in an immediately interpretable way, but the model with MFCCs allows us to resynthesise \textit{bath} vowels along the /\ae/ to /\textipa{A}/ spectrum. MFCCs and formants therefore have different strengths and limitations for analysis, depending on the goal. In this paper we demonstrate classifiers using both representations. 

Regardless of whether we work with vowel formants or MFCCs, we can view the chosen sound representation as a smooth multivariate curve over time, $X(t) \in \mathbb{R}^d$, where $t \in [0,1]$ is normalised time. In practice we assume $X(t)$ is observed with additive noise due to differences in recording devices and background noise in the recording environment. 

\section{Data sources} \label{sec:data}

In this section we describe the two data sources used in this paper.


\subsection{North-South Class Vowels} \label{sec:nscv}

The North-South Class Vowels \citep[NSCV;][]{Koshy2020_shahin} 
dataset is a collection of 400 speech recordings of the vowels /\ae/ and /\textipa{A}/ that distinguish stereotypical Northern and Southern accents in the \textit{bath} lexical set. The vowels were spoken by a group of 4 native English speakers (100 recordings per speaker).
It was collected in order to have a high-quality labelled dataset of the /\ae/ and /\textipa{A}/ vowel sounds in \textit{bath} words.
The NSCV dataset was collected with ethical approval from the Biomedical and Scientific Research Ethics Committee of the University of Warwick.

The speech recordings were collected in an experimental setup. The speakers were two male and two female adults between the ages of 18 and 55.  In order to participate they were required to be native English speakers but were not required to be proficient in Southern and Northern accents. 
To access the vowels, they were shown audio recordings as pronunciation guides, and example rhyming words such as `cat' for the /\ae/ vowel and `father' for the /\textipa{A}/ vowel. They were allowed to practice using the two vowels in the list of words, before being recorded saying a list of words using both vowels. The words were \textit{class, grass, last, fast}, and \textit{pass}. Each word was repeated 5 times using each vowel, by each speaker. The speech was simultaneously recorded with two different microphones.

The purpose of this dataset is to demonstrate a method of training accent classification models. By using vowels as a proxy for accent, it allows us to train models to distinguish between Northern and Southern accents, to the extent that they differ by this vowel. Using two microphones and having the same speaker producing both vowels allows us to train models that are robust to microphone and speaker effects. Despite the small number of speakers in this dataset, we are still able to classify vowels with high accuracy and resynthesise vowels well. A limitation of the dataset is that the speakers were not required to be native speakers of both Northern and Southern accents or have any phonetic training.

\subsection{British National Corpus} \label{sec:bnc}

The audio edition of the British National Corpus (BNC) is a collection of recordings taken across the UK in the mid 1990s, now publicly available for research \citep{BNC}. A wide range of people had their speech recorded as they went about their daily activities, and the audio recordings were annotated (transcriptions of the conversations, with information about the speakers). From this corpus we analyse utterances of the following words from the \textit{bath} lexical set, which we call the ``class'' words: \textit{class, glass, grass, past, last, brass, blast, ask, cast, fast}, and \textit{pass}.

\begin{figure}[t]
    \centering
    \includegraphics[width=.4\linewidth]{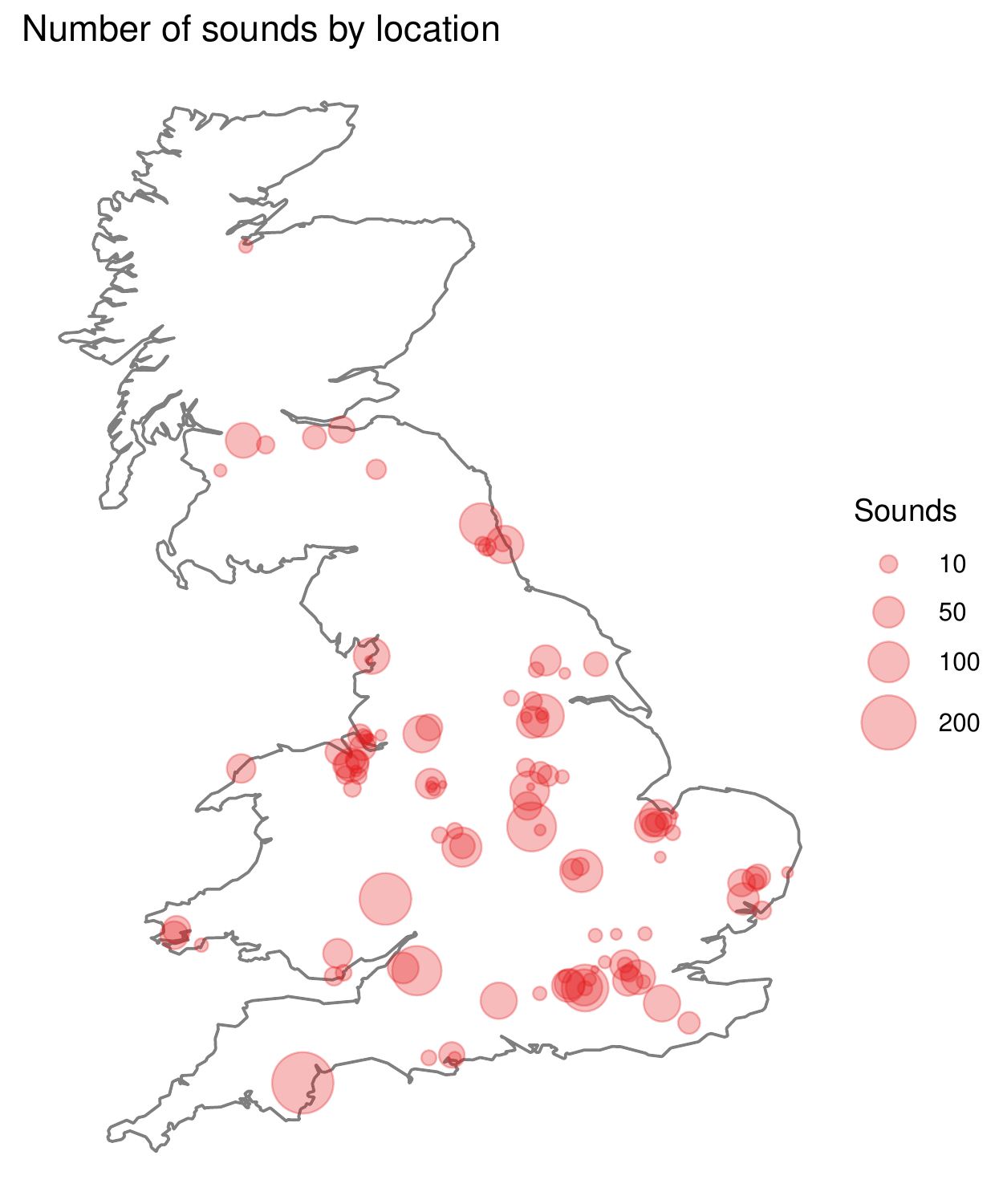}
    \includegraphics[width=.4\linewidth]{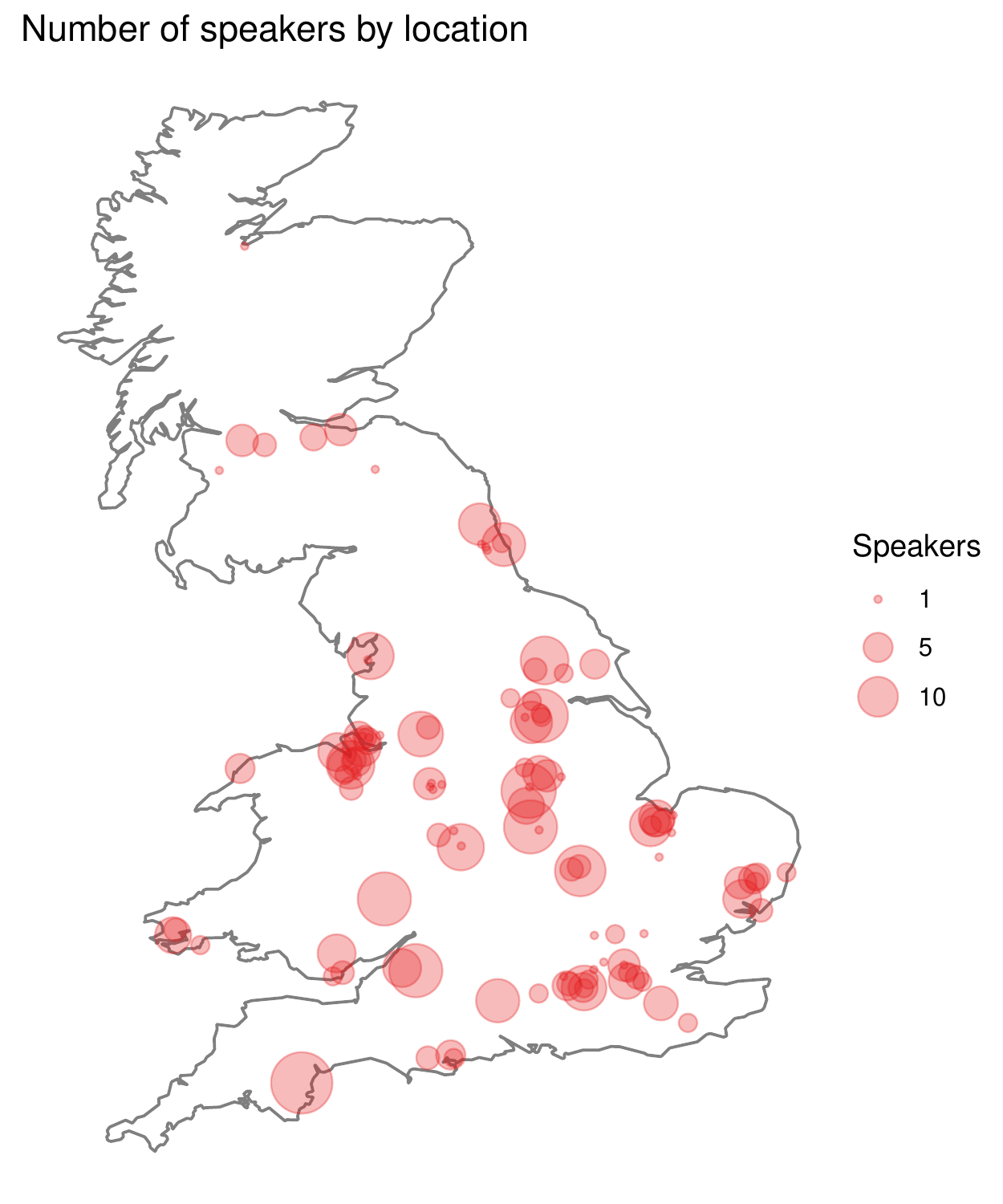}
    \caption{Each bubble is centred at a location at which we have observations in the BNC, and its size corresponds to the number of recordings (left plot) and number of speakers (right plot) at each location.}
    \label{fig:obsnum}
\end{figure}

Among the sound segments in the BNC labelled as a ``class'' word, not all of them do correspond to a true utterance of a ``class'' word by a British speaker\hl[]{, and some are not of good quality}. Some sounds were removed from the dataset using the procedure described in Appendix~\ref{app:exploration}. 
The resulting dataset contains 3852 recordings from 529 speakers in 124 locations across England, Scotland and Wales. Figure~\ref{fig:obsnum} shows the number of sounds and speakers at each location. Some speakers were recorded at multiple locations, but 94\% of them have all their recording locations within a 10 kilometre radius. 88\% of all speakers only have one recording location in this dataset.

This dataset captures a wide range of geographical locations and socio-economic characteristics, and speakers were recorded in their natural environment. It has, however, some limitations for our analysis. For example, we do not know the true origin of a speaker, so unless the metadata shows otherwise, we must assume that speakers' accents are representative of the location where they were recorded. There are very few speech recordings available from the North, especially Scotland. The timestamps used to identify word boundaries are often inaccurate, and the sound quality varies widely between recordings, due to background noise and the different recording devices used.

\subsection{Transforming sounds into data objects} \label{sec:preprocess-steps}


Each speech recording in the BNC and NSCV datasets was stored as a mono-channel 16 kHz \texttt{.wav} file. The raw formants were computed using the \texttt{wrassp} R package \citep{wrassp}. At each single time point the first four formants were computed, and this is done at 200 points per second. A sound of length 1 second is thus represented as a $200\times4$ matrix, where each column corresponds to one formant curve. For each vowel sound, raw MFCCs were extracted using the \texttt{ahocoder} software \citep{Erro2011, Erro2014}, which also computes them at 200 points per second. Hence a sound of length 1 second would be represented as a $200\times40$ matrix, where each column represents one MFCC curve. 

\begin{figure}[h]
    \centering
    \includegraphics[width=\textwidth]{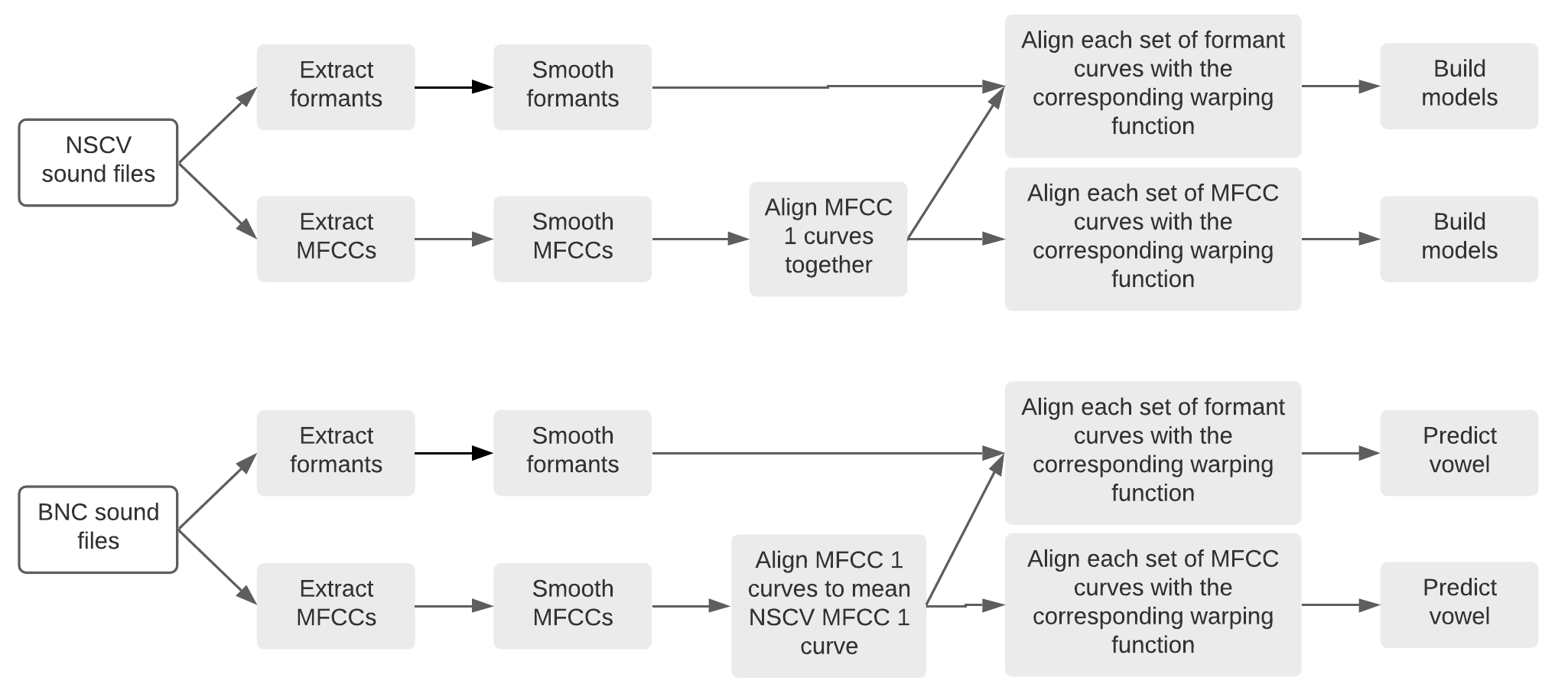}
    \caption{Summary of preprocessing steps.}
    \label{fig:flowchart}
\end{figure}

We smooth the raw formants and raw MFCCs in order to remove unwanted variation due to noise, and to renormalise the length of the curves by evaluating each smoothed curve at a fixed number of time points \citep{Ramsay2005}. 

Assuming a signal plus noise model on the raw formants and raw MFCCs, we smooth and resample them on an equidistant grid of length $T=40$. Since the raw formants exhibit large jumps that are physiologically implausible, we smooth them using robust loess \citep[R function \texttt{loess}][]{Cleveland1979} with smoothing parameter $l=0.4$ and using locally linear regression. The raw MFCCs are less rough, and we smooth them using cubic splines \citep[R function \texttt{smooth.spline},][]{R2020}  with knots chosen at each point on the time grid and smoothing parameter chosen by cross-validation for each curve. We have used $T=40$ in this analysis because it captures the main features while not inflating the dataset too much. We do not model vowel duration, which also depends on other factors, such as speech context \citep{Clayards}. Other implementations and smoothing methods could be used here, such as the R package \texttt{mgcv} for smoothing MFCCs with cubic splines, and robust smoothing for formants using the scaled t family.



Finally, we perform an alignment step to reduce misalignments between NSCV curves and BNC curves. This is necessary because the BNC speech recordings often have inaccurate timestamps and this can cause their vowels to be misaligned with the NSCV curves. Since we classify BNC vowels using models trained on NSCV curves, these misalignments can cause inaccuracies in the predictions. We consider the differences in relative timing of the vowel in the sound to be due to a random phase variation; alignment or registration of curves allows us to reduce the effect of this phase variation \citep{Ramsay2005}. We use the approach of \citet{Srivastava2011}, where the Fisher--Rao metric distance between two curves is minimised by applying a nonlinear warping function to one of the curves. 

The first MFCC curve (MFCC 1) of each sound contains the volume dynamics. To align NSCV vowels, we first align all NSCV MFCC 1 curves together. These warping functions are then applied to the formant curves and other MFCC curves from the same vowels, since they come from the same underlying sounds. For each BNC vowel, we first align its MFCC 1 curve to the mean aligned NSCV MFCC 1 curve, and then use the obtained warping function to align all the other MFCC curves and formant curves from the same vowel. Alignment was performed using the R package \texttt{fdasrvf} \citep{fdasrvf2020}, and the preprocessing steps are summarised in Figure~\ref{fig:flowchart}.

\section{Classifying accents} \label{sec:classify}

In this section, we will present two models for classifying \textit{bath} vowels as Southern or Northern.  

\subsection{Modeling formants} \label{sec:formant-model}

Our first task is to build a classifier to classify \textit{bath} vowels as Northern or Southern. The model uses the fact that the first two formants $\text{F}_1$ and $\text{F}_2$ are known to predominantly differentiate vowels, and higher formants do not play as significant a role in discriminating them \citep{Adank, Johnson2005}. It has been suggested that the entire trajectory of formants are informative even for stationary vowels like the \textit{bath} vowels, and they should not be considered as static points in the formant space \citep{Johnson2005}; (see also the discussion in Section \ref{sec:discussion}). This suggests the use of formant curves as functional covariates when modelling the vowel sounds. 
Since the dynamic changes in the formant curves are not drastic, we do not believe there are time-localised effects of the formants, so we use the entire formant curve as a covariate with a roughness penalty. Due to the nested structure of the NSCV corpus with 100 speech recordings from each speaker, we also include a random effect term to account for variation between speakers.


Now we can propose the following functional logistic regression model to classify accents:
\begin{equation}
    \logit(p_{ij}) = \beta_0 + \int_{0}^{1}\text{F}_{2ij}(t)\beta_1(t)dt + \gamma_j, \label{eq:loggam}
\end{equation}
%
\begin{figure}[t]
    \centering
    \includegraphics[width=4in]{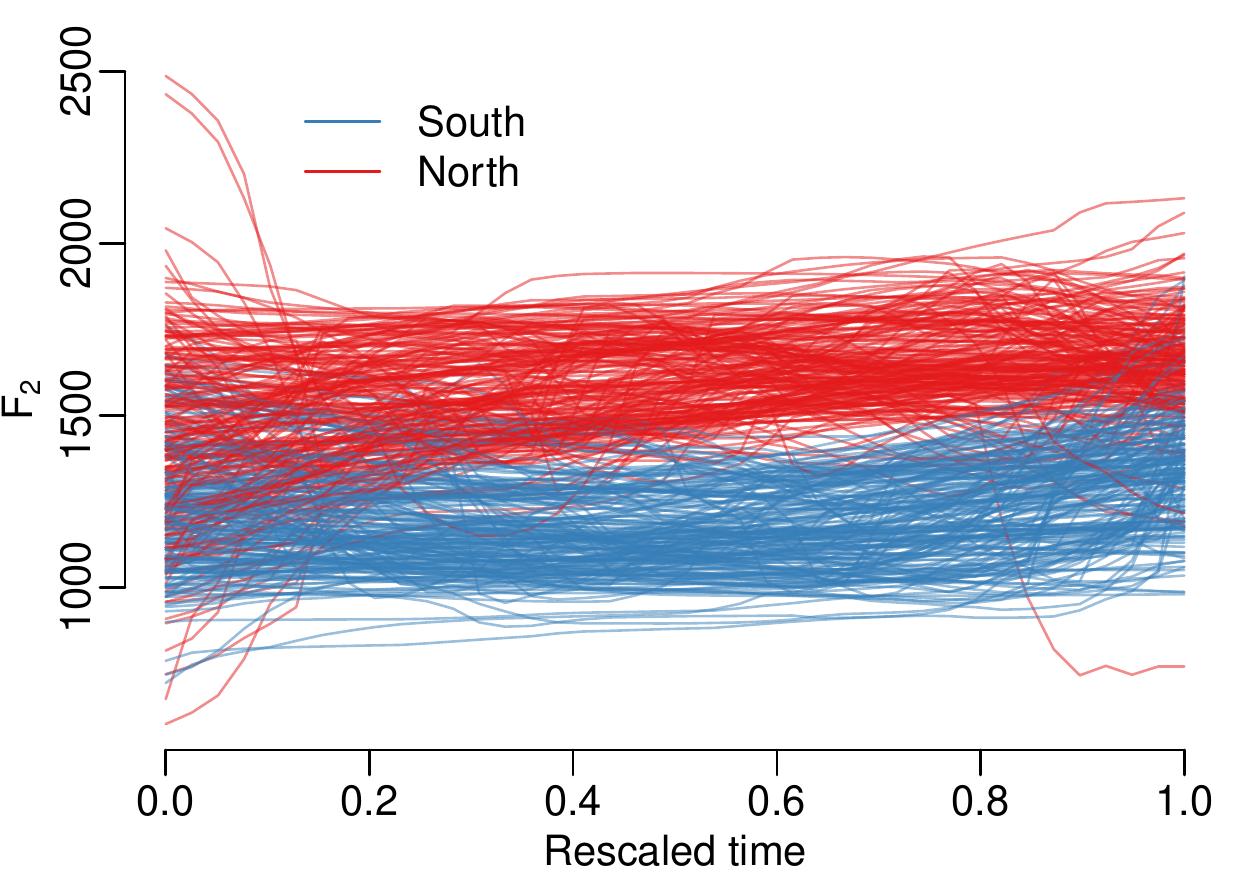}
    \caption{Smoothed and aligned $\text{F}_2$ formant curves for the NSCV vowels. Each curve corresponds to one vowel sound.}
    \label{fig:f2}
\end{figure}
where $p_{ij}$ is the probability of sound $i$ from speaker $j$ being Southern, $\text{F}_{2ij}(t)$ is the value of the $\text{F}_2$ curve at time $t$ for sound $i$ from speaker $j$, and $\gamma_j \sim N(0, \sigma_s^2)$ is a random effect for speaker $j$. The functional covariate contributes to the predictor through a linear functional term. The integral is from 0 to 1 since we have normalised the length of all sounds during preprocessing. The function $\beta_1(t)$ is represented with a cubic spline with knots at each time point on the grid, and its ``wiggliness'' is controlled by penalising its second derivative. Model selection was done by comparing the adjusted AIC \citep{Wood} to decide which other terms should be included in the model. Further details from the model selection procedure are given in Appendix \ref{app:modelselection}, where we also consider simpler non-functional models. The model was fitted using the \texttt{mgcv} package in R \citep{Wood2011}.

\begin{figure}[h]
    \centering
    \includegraphics[width=4in]{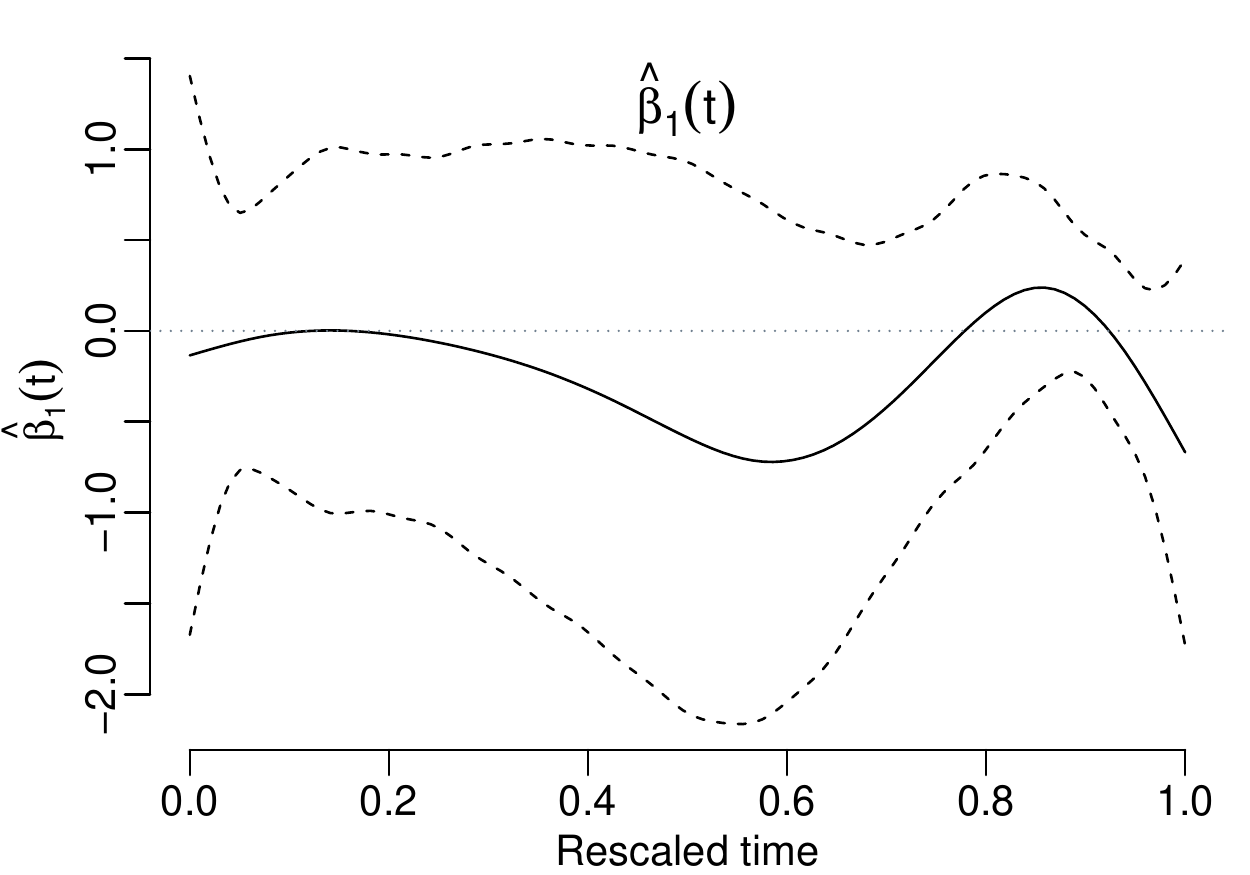}
    \caption{$\hat{\beta_{1}}(t)$ shows that a lower $\text{F}_2$ region towards the middle of the sound indicates a more Northern vowel sound. The dashed lines are 95\% pointwise confidence intervals of the coefficient curve.}
    \label{fig:betahatt}
\end{figure}

The fitted coefficient curve $\hat \beta_1(t)$, shown in Figure~\ref{fig:betahatt}, reveals that middle section of the $\text{F}_2$ curve is important in distinguishing the vowels. A lower $\text{F}_2$ curve in this region indicates a Northern /\ae/ vowel. From a speech production perspective, this corresponds to the Northern vowel being more ``front'', which indicates that the highest point of the tongue is closer to the front of the mouth, compared to the Southern vowel. 
The point estimate for $\beta_0$ is 328.0 (p-value $= 0.267$, 95\% CI $[-250.85, 906.87]$). The variance component explained by the speaker random effects is $\hat{\sigma}_s^2 = 0.006$ (p-value $= 0.776$). 


\begin{figure}[h]
    \centering
    \includegraphics{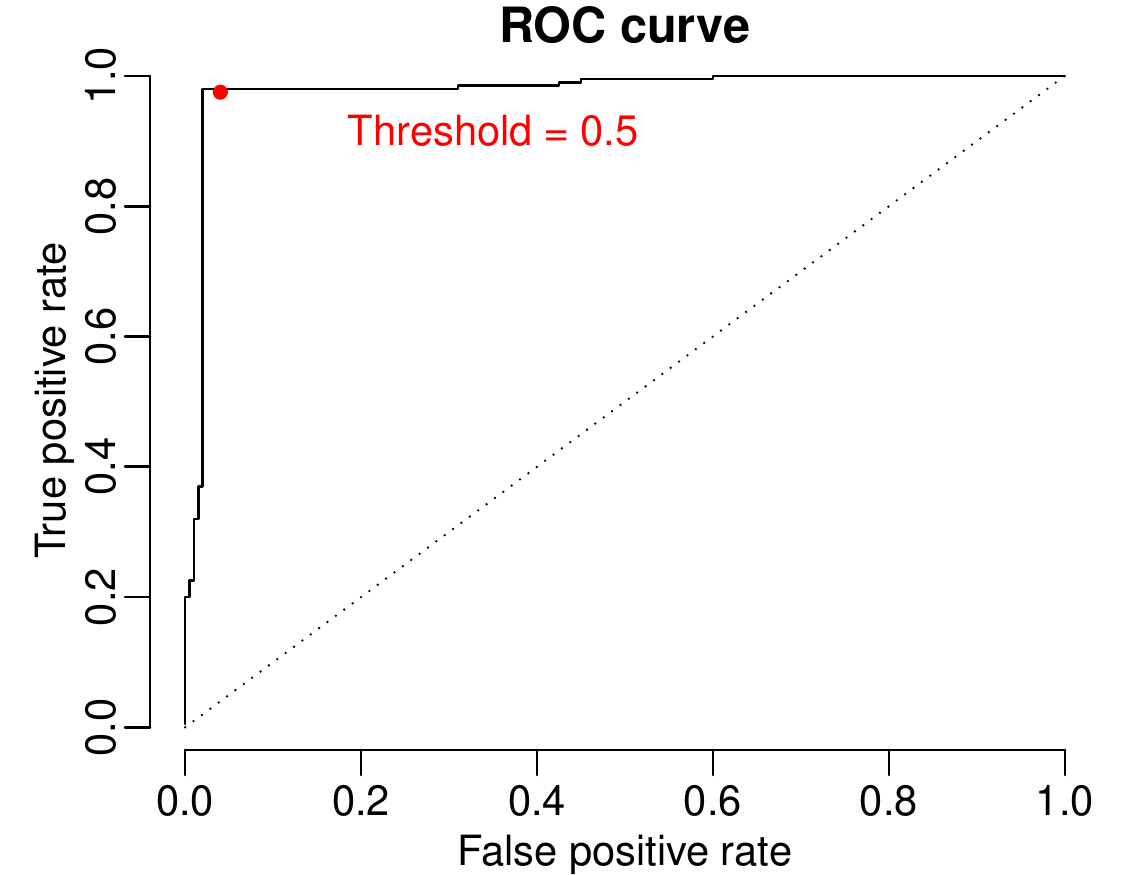}
    \caption{ROC curve for the functional logistic regression model. The dotted line corresponds to random guessing and the red dot corresponds to using a threshold of 0.5 to classify vowels.}
    \label{fig:roc}
\end{figure}

This model assigns a ``probability of being Southern'' to a given vowel sound, by first aligning the sound to the mean NSCV sound using MFCC 1, and then plugging its formants into \eqref{eq:loggam}. We classify a vowel sound as Southern if its predicted probability of being Southern is higher than $0.5$. 
We can estimate the classification accuracy of this model through cross-validation. The model was cross-validated by training it on 3 speakers and testing on the fourth speaker's vowels, and repeating this 4 times by holding out each speaker in the dataset. Using a random split of the data instead would lead to overestimated accuracy, because different utterances by the same speaker cannot be considered independent. The cross-validated accuracy is 96.75\%, and the corresponding confusion matrix is shown in Table~\ref{table:conf}. We can also compare the performance of this model for different classification thresholds, using the ROC curve in Figure~\ref{fig:roc}.

\begin{table}[h]
    \caption{Cross-validated confusion matrix for the functional logistic regression.}
    \label{table:conf}
    \centering
    \begin{tabular}{rcrr}
        \toprule
        {} &\phantom{} &\multicolumn{2}{c}{\textbf{Truth}}\\
        \cmidrule{3-4} 
                        && North & South \\ 
        \textbf{Prediction} \\
         North && 196 & 4 \\
        South  && 4 & 196\\
        \bottomrule
    \end{tabular}

\end{table}

\subsection{Modeling MFCCs} \label{sec:mfcc-model}

We will now present another approach to classifying vowel sounds, which uses the MFCC curves obtained from each vowel recording. We have 40 smoothed MFCC curves for each sound. 
Unlike with formants, we do not have prior knowledge about which curves contain information about the vowel quality. Additionally, since MFCC curves contain information about different parts of the frequency spectrum, they are not independent and the co-variation between curves is important. For example, setting an MFCC curve (or a region of the curve) to a constant value distorts the resulting sound. Hence a multivariate functional regression approach with $\ell_1$ penalty to remove certain curves from the model would not be appropriate, and we need to incorporate information from potentially all the MFCC curves in our model. The problem of concurvity between MFCC curves can also destabilise the resulting coefficient curve estimates in such an approach. 

Interpreting the shapes of the curves is also not as useful here since MFCC trajectories do not have a physical interpretation as formants do. We are more interested in the model's ability to resynthesise vowels by capturing as much relevant information about vowel quality as possible. Hence we use functional principal components analysis to capture the co-variation of the MFCC curves. This step essentially generates new features by reparametrising the MFCC curves, which we can then use to fit the classification model.
We select the most informative functional principal components to be in the model through $\ell_1$ penalisation.


\subsubsection{Functional Principal Component Analysis}

Functional principal component analysis \citep[FPCA;][]{Ramsay2005} is an unsupervised learning technique which identifies  the different modes of variation in a set of observed smooth curves $\{X_i: [0,1] \rightarrow \mathbb{R},\, i = 1 ,\ldots, n\}$. It is very similar to standard principal component analysis, except that the variables are curves instead of scalar features, and each functional principal component (FPC) is also a curve instead of a vector.  

Assuming that the curves $\{X_i\}$ are centred, the $k$th FPC is a smooth curve $\varphi_k:[0,1] \rightarrow \mathbb{R}$ which maximises
\[
\frac{1}{n} \sum_{i=1}^{n} \left( \int \varphi_k(t) X_i(t) dt \right) ^2,
\]
subject to $\int \varphi_k(t)^2 dt = 1$ and $\int\varphi_k(t)\varphi_j(t)dt = 0$ for all $j < k$; there is no constraint for $k=1$. The functional principal component score (FPC score) of curve $i$ with respect to principal component $\varphi_k$ is $s_{ik} = \int \varphi_k(t)X_i(t)dt$. 

In multivariate FPCA, each observation is a curve in $\mathbb{R}^M$, and the set of observations is $\{ {\boldsymbol X}_i=(X_i^{(1)}, X_i^{(2)}, \ldots, X_i^{(M)}): [0,1] \rightarrow \mathbb{R}^M,\, i = 1 ,\ldots, n\}$. Amongst the existing variants of multivariate FPCA \citep{chiou2014multivariate,Happ2018}, we use the following one:
assuming that the curves $\{\boldsymbol X_i\}$ are centred, the $k$th FPC is a smooth multivariate curve, defined as ${\boldsymbol \varphi}_k = (\varphi_k^{(1)}, \varphi_k^{(2)}, \ldots, \varphi_k^{(M)}):[0,1] \rightarrow \mathbb{R}^M$ which maximises
\[
\frac{1}{n} \sum_{i=1}^{n} \sum_{j=1}^{M} \left( \int \varphi_k^{(j)}(t) X_i^{(j)}(t) dt\right)^2
\]
subject to $\sum_{j=1}^M \int [ \varphi_k^{(j)}(t) ]^2 dt = 1$ and $\sum_{j=1}^{M} \int\varphi_k^{(j)}(t)\varphi_l^{(j)}(t)dt = 0$ for all $l < k$. The $k$-th FPC score of ${\boldsymbol X}_i$ is defined as $s_{ik} = \sum_{j=1}^M \int \varphi_k^{(j)}(t) X_i^{(j)}(t) dt$.

In our case, the curves $\{ {\boldsymbol X}_i\}$ are the MFCC curves with $M=40$. Each curve $\boldsymbol{X_i}$ discretised on a grid of $T$ equally spaced time points, yielding a $T \times M$ matrix, which is then transformed by stacking the rows into a vector in $\mathbb{R}^{MT}$. The whole dataset is then represented as an $n \times MT$ matrix, which contains observations as rows. The (discretised) FPCs and their scores can therefore be directly computed using a standard implementation of (non-functional) PCA, such as \texttt{prcomp} in R \citep{R2020}. 

\begin{figure}[h]
    \centering
    \includegraphics[width=4.5in]{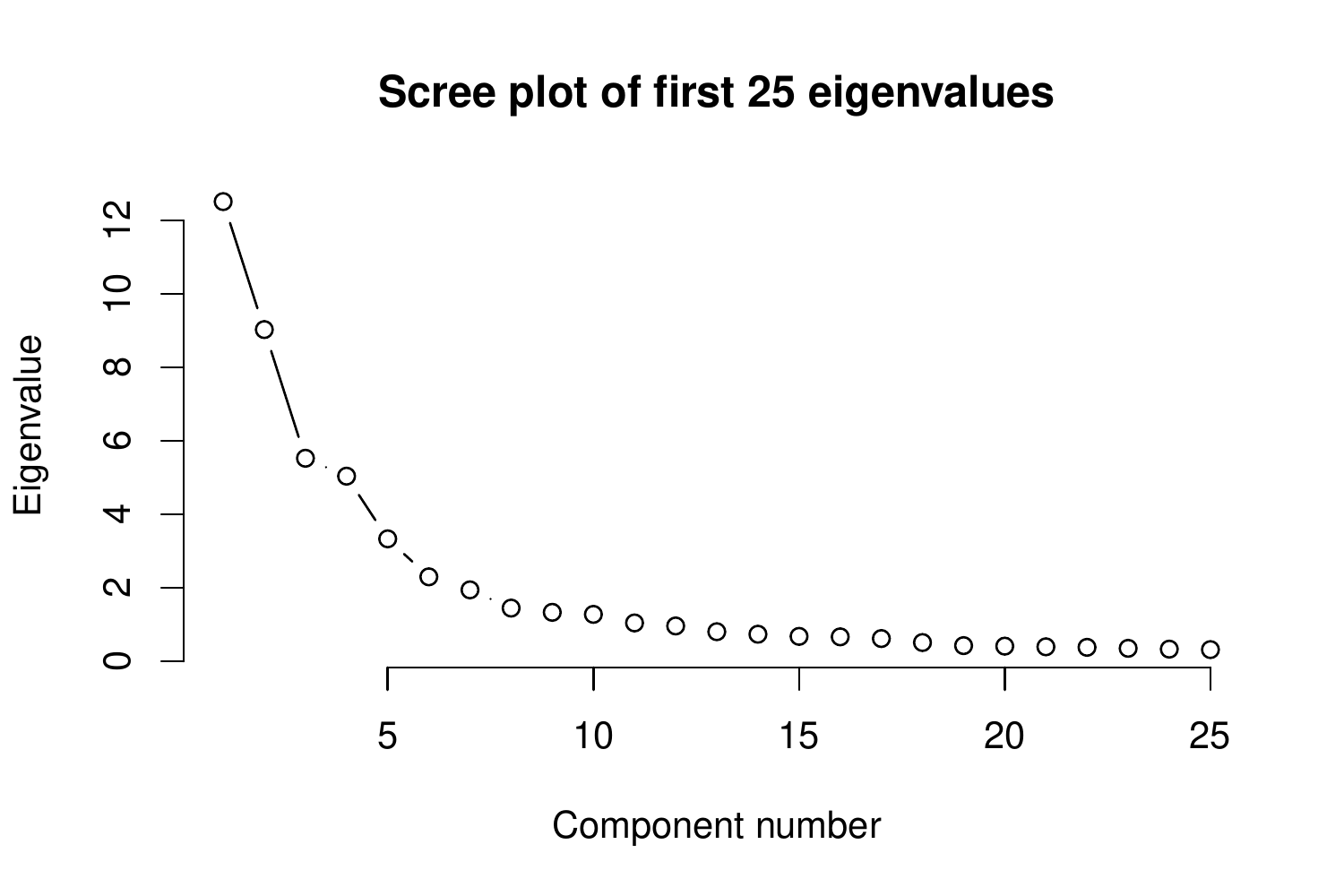}
    \caption{First 25 eigenvalues of the functional principal components of the MFCCs.}
    \label{fig:screeplot}
\end{figure}

Before performing the FPCA we centre each MFCC 1 curve at zero, because the average level of MFCC 1 mainly contains differences in the overall volume of the sound, which is influenced by factors other than accent. Centring the curve at zero retains the volume dynamics in the vowel while normalising the overall volume between sounds. Since there are 400 observations in the NSCV training data, we can decompose the MFCC curves into (at most) 400 functional principal components. The first 25 eigenvalues of the FPCs obtained are plotted in Figure~\ref{fig:screeplot}.

\subsubsection{$\ell_1$-Penalised Logistic Regression}

$\ell_1$-penalised logistic regression \citep[PLR;][]{Hastie2017} can be used for binary classification problems when we have many covariates (here we have $p=400$ FPC scores which we could include in the model, which corresponds to a reparametrisation of the MFCC curves without losing any information). Through the penalisation and model fitting procedure, a smaller subset of covariates are chosen in the final model.

The model is the same as for the usual logistic regression: if $Y$ is a Bernoulli random variable and $\boldsymbol{X} \in \mathbb{R}^p$ is its covariate vector, the model is
\[
    \logit( \mathbb{P}(Y = 1 | \boldsymbol{X} = \boldsymbol{x}) ) = \beta_0 + \boldsymbol{\beta}^\mathsf{T} \boldsymbol{x},
\]
but it is fitted with an added $\ell_1$ penalty on the regression coefficients to deal with high-dimensionality, which encourages sparsity and yields a parsimonious model.
In our setting, 
if $y_i = 1$ if sound $i$ is Southern, $y_i=0$ if it is Northern, and $\boldsymbol{x}_i \in \mathbb{R}^{400}$ is a vector of its 400 FPC scores, PLR is fitted by solving
\begin{equation}
    \label{eq:PLR}
    (\hat{\beta_0}, \hat{\boldsymbol \beta}) = \arg \max_{\beta_0, {\boldsymbol \beta}} \sum_{i=1}^n \left(y_i (\beta_0 + {\boldsymbol \beta}^\mathsf{T} {\boldsymbol x}_i) - \log(1 + e^{\beta_0+ {\boldsymbol \beta}^\mathsf{T} {\boldsymbol x}_i})\right) - \lambda \sum_{j=1}^{p} \lvert \beta_j \rvert,
\end{equation}
where $\lambda \geq 0$ is a penalty weight. 
Notice that the first term in \eqref{eq:PLR} is the usual log-likelihood, and the second term is an $\ell_1$ penalty term. The penalty $\lambda$ is chosen by 10-fold cross-validation.
A new sound with FPC scores vector $\boldsymbol{x_*}$ is assigned a ``probability of being Southern'' of $\ilogit( \hat \beta_0 + \boldsymbol{\hat \beta}^\mathsf{T} \boldsymbol{x_*} )$, where 
$\ilogit(\cdot)$ is the inverse logit function. We classify the sound as Southern if $\ilogit( \hat \beta_0 + \boldsymbol{\hat \beta}^\mathsf{T} \boldsymbol{x_*} ) \geq 0.5$.
 
\begin{figure}[h]
    \centering
    \includegraphics[width=4in]{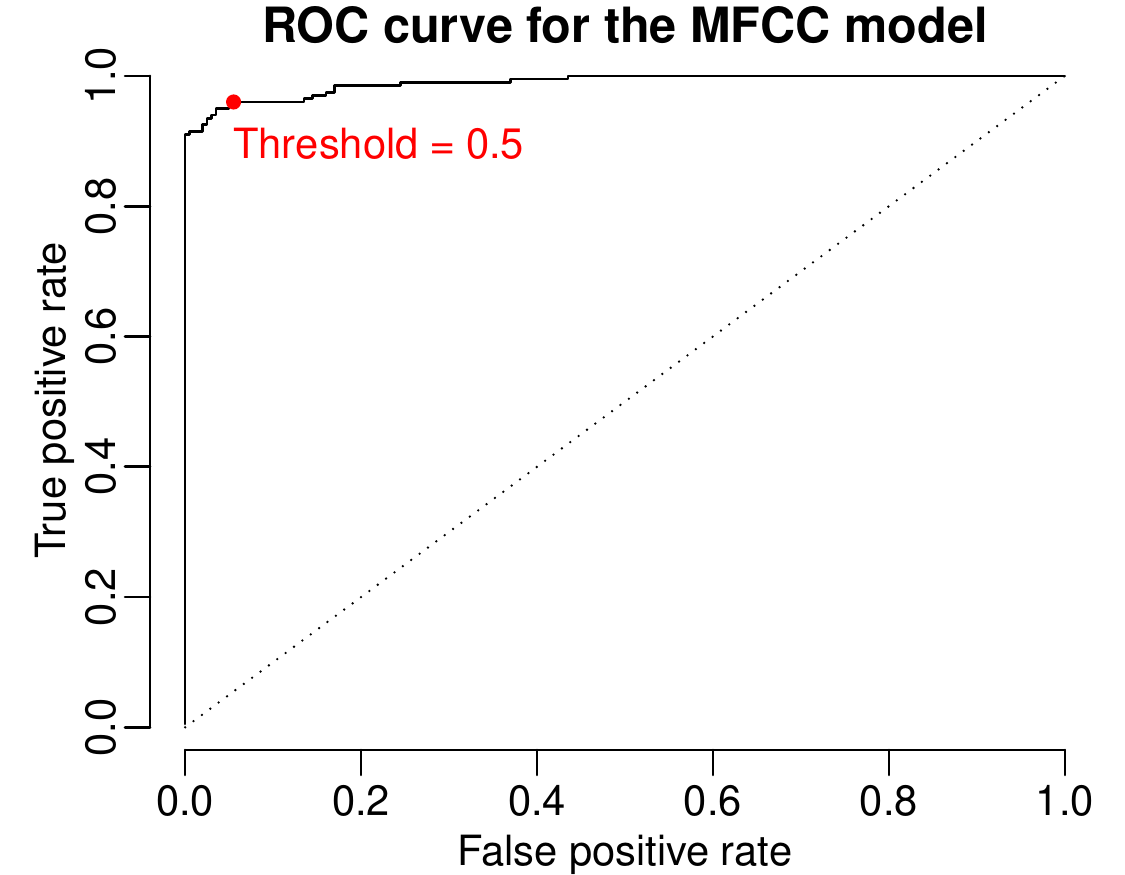}
    \caption{ROC curve for the MFCC model using penalised logistic regression classifier.}
    \label{fig:plr_roc}
\end{figure}

We can estimate the accuracy of the model by cross-validating using individual speakers as folds, as in the functional linear model of Section~\ref{sec:formant-model}. Within each training set, we first perform the FPCA to obtain the FPCs and their scores. Then we cross-validate the penalised logistic regression model to find the optimal penalty $\lambda$, and retrain on the whole training set with this $\lambda$. Finally, we project the test speaker's sounds onto the FPCs from the training set to obtain the test FPC scores, and use them to classify the vowel of each sound using the predicted probabilities from the trained model. This process is repeated holding out each speaker in turn. The cross-validated accuracy of this model is 95.25\%. The confusion matrix is shown in Table~\ref{table:plrconf}, and the ROC curve is shown in Figure~\ref{fig:plr_roc}. 

\begin{figure}[h]
    \centering
    \includegraphics[width=4in]{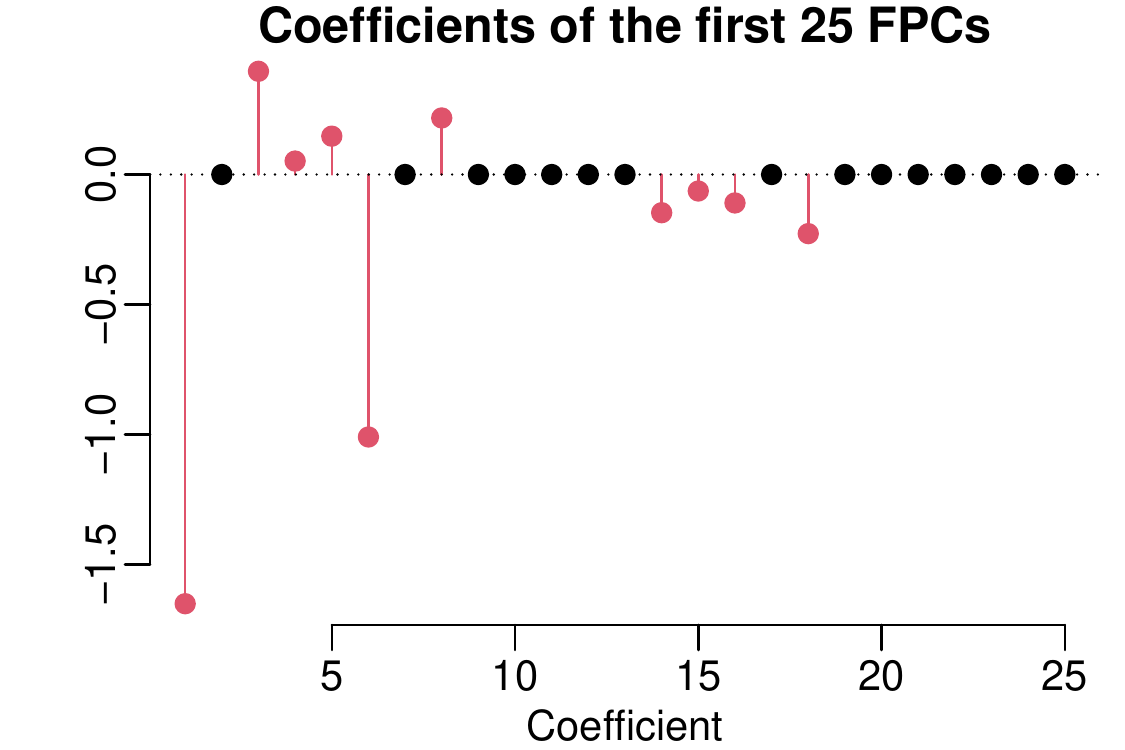}
    \caption{The first 25 entries of $\hat{\boldsymbol \beta}$ maximising \eqref{eq:PLR}. Nonzero entries are shown in red. All the later entries are zero, not shown here.}
    \label{fig:plr_coefs}
\end{figure}

To fit the full model, we use the entire dataset to cross-validate to choose the best $\lambda$, and then refit on the entire dataset using this penalty. The entries of $\bf \beta$ are essentially weights for the corresponding FPCs. By identifying the FPC scores which have nonzero coefficients, we can visualise the weighted linear combination of the corresponding FPCs which distinguish Northern and Southern vowels. In total 10 FPCs had nonzero weights, and all of the chosen FPCs were within the first 20. A plot of the first 25 coefficient values is given in Figure~\ref{fig:plr_coefs}.

\begin{table}[h]
    \caption{Cross-validated confusion matrix for the penalised logistic regression classifier.}
    \label{table:plrconf}
    \centering
    \begin{tabular}{rcrr}
        \toprule
        {} &\phantom{} &\multicolumn{2}{c}{\textbf{Truth}}\\
        \cmidrule{3-4} && North & South \\
        \textbf{Prediction} \\
        North && 189 & 8   \\
        South && 11  & 192 \\
        \bottomrule
    \end{tabular}
\end{table}


\subsubsection{Resynthesising vowel sounds}
\label{sec:resynthesising}

\begin{figure}[h]
    \centering
    \includegraphics[width=4.5in]{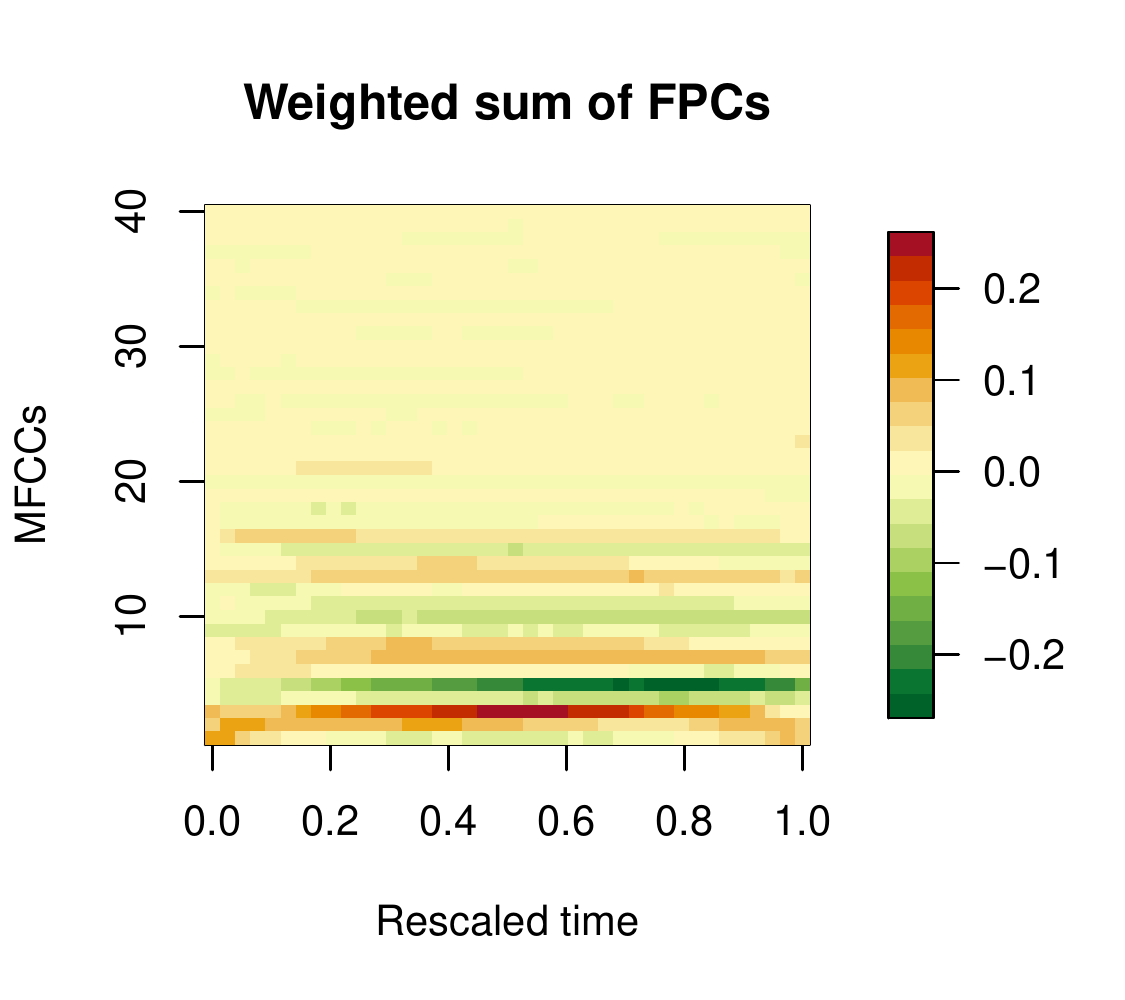}
    \caption{This image shows the 
    MFCCs of \eqref{eq:mfcc_making_more_southern} which make a vowel sound more Southern. Each row of the image is an MFCC curve.}
    \label{fig:contrib}
\end{figure}

\begin{figure}[h]
    \centering
    \includegraphics[width=\textwidth]{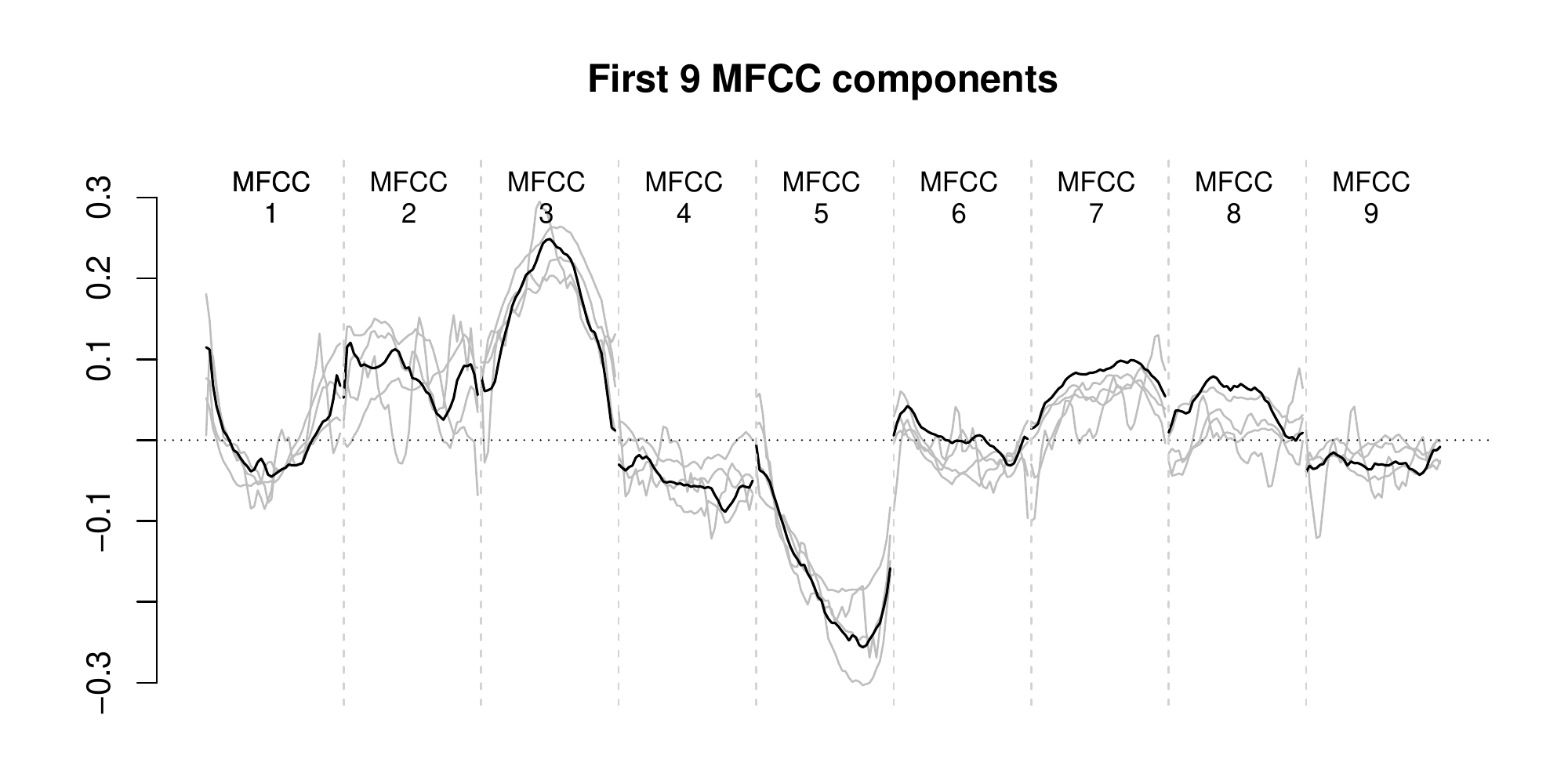}
    \caption{The first 9 MFCCs from model \eqref{eq:mfcc_making_more_southern}, which correspond to the bottom 9 rows of the matrix in Figure~\ref{fig:contrib}, plotted sequentially. We can see that MFCC 3 and 5 have large contributions. The grey lines are the MFCC curves obtained in each cross-validation fold and thicker black lines are from the final model.}
    \label{fig:contrib_first_9}
\end{figure}

The combined effect of the functional principal components that are predictive of accent is given by the function 
\begin{equation}
    \label{eq:mfcc_making_more_southern}
    \sum_{k=1}^{400} \hat{\beta}_{k} \hat {\boldsymbol \varphi_k}: [0,1] \to \mathbb{R}^{40}.
\end{equation}
Discretising this function on an equispaced grid of $T$ points yields a $T \times 40$ matrix, which can be visualised (Figure~\ref{fig:contrib}), or interpreted as a set of MFCC curves (Figure~\ref{fig:contrib_first_9}).

This MFCC matrix captures the difference between the /\ae/ and /\textipa{A}/ vowels. Since MFCCs can be used to synthesise speech sounds, we can now make a given \textit{bath} vowel sound more Southern or Northern, through the following procedure:
We first extract the MFCCs for the entire utterance of a \textit{bath} word, as a $T \times 40$ matrix where $T$ is determined by the length of the sound. With manually identified timestamps we find the $T_v$ rows of this matrix which correspond to the vowel in the word. We align MFCC 1 of this vowel to the mean NSCV MFCC 1 curve, to obtain the optimal warping function for the sound. The MFCC matrix in Figure~\ref{fig:contrib} is `unwarped' using the inverse of this warping function, resampled at $T_v$ equidistant time points, and padded with $T - T_v$ rows of zeroes corresponding to the rest of the sound's MFCCs (which we do not change). We can then add multiples of this $T\times40$ matrix to the original sound's MFCC matrix and synthesise the resulting sounds using \texttt{ahodecoder} \citep{Erro2014}. Adding positive multiples of the matrix makes the vowel sound more Southern, while subtracting multiples makes it sound more Northern. In the supplementary material we provide audio files with examples of this: \texttt{blast-StoN.wav} contains the word ``blast'' uttered in a Southern accent and perturbed towards a Northern accent, and \texttt{class-NtoS.wav} contains the word ``class'' uttered in a Northern accent and perturbed towards a Southern accent. Both of these original vowels were new recordings, and not from the NSCV corpus.


\section{Modelling geographic variation} \label{sec:maps}

In this section we demonstrate an approach for visualising the trap--bath split by combining data from the BNC with the trained accent classifiers described in Sections~\ref{sec:formant-model} and~\ref{sec:mfcc-model}. For each BNC speaker we predict the probability of their vowel sound being Southern (using in turn the formant model and the MFCC model), and then smooth the predicted probabilities spatially using a soap film smoother.

The BNC \textit{bath} vowels contain more variation than the NSCV dataset. This is partly because of more natural variation in conversational speech, as well as other factors like poor quality of some recordings and background noise. The BNC recordings also contain whole words and not only the vowel portion of the utterance. The timestamps for word boundaries are often inaccurate and many sounds are either a partial word, or contain parts of other words or speech from other speakers. It is hard to automatically detect the vowel portions within these recordings. We address this issue through the alignment step described in Section \ref{sec:preprocessing} to align each sound to the NSCV using the mean NSCV MFCC 1 curve.

A single representative sound can be constructed for each speaker by taking an average of these aligned formant and MFCC curves from the speaker's utterances. By resynthesising the sound of the average MFCC curves, we can hear that it retains the quality of a \textit{bath} vowel, so we use these average MFCCs and formants as representative of each speaker's vowel sound. For each speaker we obtain two predicted probabilities of their accent being Southern (one based on the formants, and one on the MFCCs), using models of Sections~\ref{sec:formant-model} and \ref{sec:mfcc-model}.  Notice that for each speaker, plugging this average sound's formants (MFCCs) into the trained models of Sections~\ref{sec:formant-model} (Section~\ref{sec:mfcc-model}) yields the same predicted logit probability as if we averaged the logit probabilities from each sound's aligned formants (aligned MFCCs). The averaging step used to get speaker-specific probabilities ensures that the model is not unduly influenced by individual speakers who have many recordings at one location, while also reducing the predicted probability uncertainties. Where a speaker has recordings at multiple locations, we attribute their average sound to the location with most recordings.

At each location $(\texttt{lon}, \texttt{lat})$ in Great Britain, we denote by $f(\texttt{lon}, \texttt{lat})$ the logit of the expected probability of a randomly chosen person's accent being Southern. 
We will estimate this surface using a spatial Beta regression model:
%
\begin{eqnarray}
    \label{eq:betaReg}
    p_{ij} &\stackrel{\text{iid}}{\sim}& \text{Beta}(\mu_i \nu, \nu (1-\mu_i)), \quad j \in \{1, \ldots, n_i\}\\
    \logit(\mu_i) &=& f(\texttt{lon}_i, \texttt{lat}_i), \nonumber
\end{eqnarray}
where  $p_{ij} \in [0,1]$ is the predicted probability of the $j$-th speaker's accent at location  $(\texttt{lon}_i, \texttt{lat}_i)$ being Southern, $j=1,\ldots, n_i$.
The surface $f$ is estimated using a soap film smoother within the geographic boundary of Great Britain. 
A single value of $\nu > 0$ is estimated for all observations, as in GLMs. Notice that $\ilogit(f(\texttt{lon}_i,\texttt{lat}_i)) = \mu_i = \mathbb{E}(p_{ij})  \in [0,1]$ represents the expected probability of the accent of a randomly chosen person being Southern at location $(\texttt{lon}_i, \texttt{lat}_i)$.  

One may instead consider fitting a linear model directly on the estimated linear predictor scores obtained from the formant or MFCC models. This linear approach would not be robust to the estimated linear predictor taking large values (which is the case with our data), even though the associated probabilities are essentially equal to one (or zero). Our approach alleviates this by smoothing predictions on the probability scale which makes it less influenced by outliers. Link functions other than the logit could also be used in alternative approaches.

Let us now recall the soap film smoother.
The soap film smoother \citep{Wood2008} is a nonparametric solution to spatial smoothing problems, which avoids smoothing across boundaries of a bounded non-convex spatial domain. 
We observe data points $\{(x_i, y_i, z_i), i=1,\ldots,n\}$, where $z_i$ are the responses with random noise and $\{(x_i, y_i)\}$ lie in a bounded region $\Omega \subset \mathbb{R}^2$. The objective is to find the function $f:\Omega \rightarrow \mathbb{R}$ which minimises
\[
\sum_{i=1}^n (z_i - f(x_i, y_i))^2 + \lambda \int_{\Omega} \left( \frac{\partial^2f}{\partial x^2} + \frac{\partial^2f}{\partial y^2}\right)^2 dx dy.
\]
The smoothing parameter $\lambda$ is chosen through cross-validation. The soap film smoother is implemented in the R package \texttt{mgcv} \citep{Wood2011}.

In our model \eqref{eq:betaReg}, the predicted Southern accent probabilities $\{p_{ij}\}$ of individual speakers are observations at different locations $\{(\texttt{lon}_i, \texttt{lat}_i)\}$ in Great Britain, and we use the soap film smoother to construct a smooth surface $f(\cdot, \cdot)$ to account for the geographic variation. We can compare the results using accent predictions from the two classification models proposed in the previous section.

\begin{figure}[h]
    \centering
    \begin{subfigure}[b]{.49\textwidth}
        \centering
        \includegraphics[trim=50 0 20 0, clip=TRUE, width=0.9\textwidth]{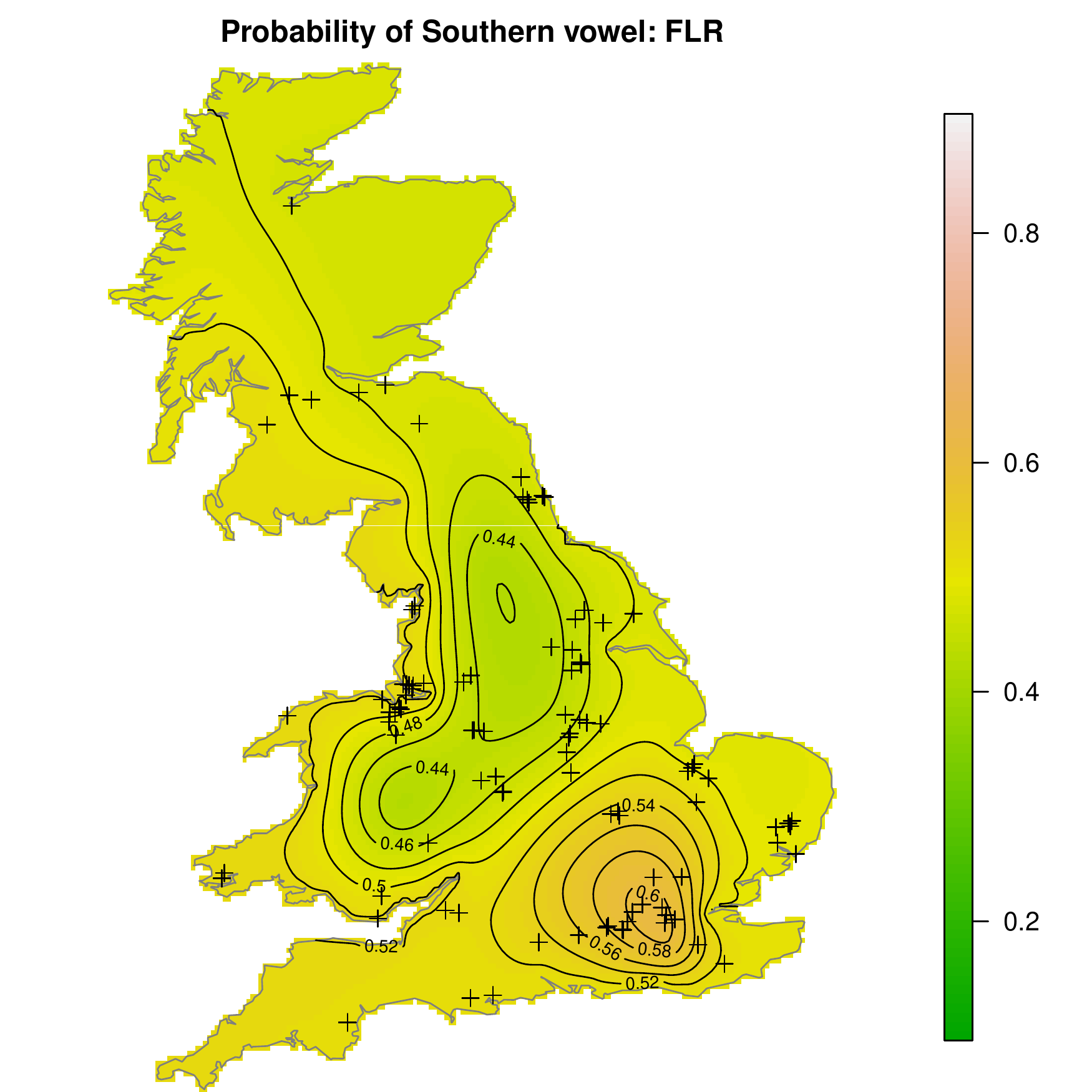}
        \caption{Map using formants.}
    \end{subfigure}
    \begin{subfigure}[b]{.49\textwidth}
        \centering
        \includegraphics[trim=50 0 20 0, clip=TRUE, width=0.9\textwidth]{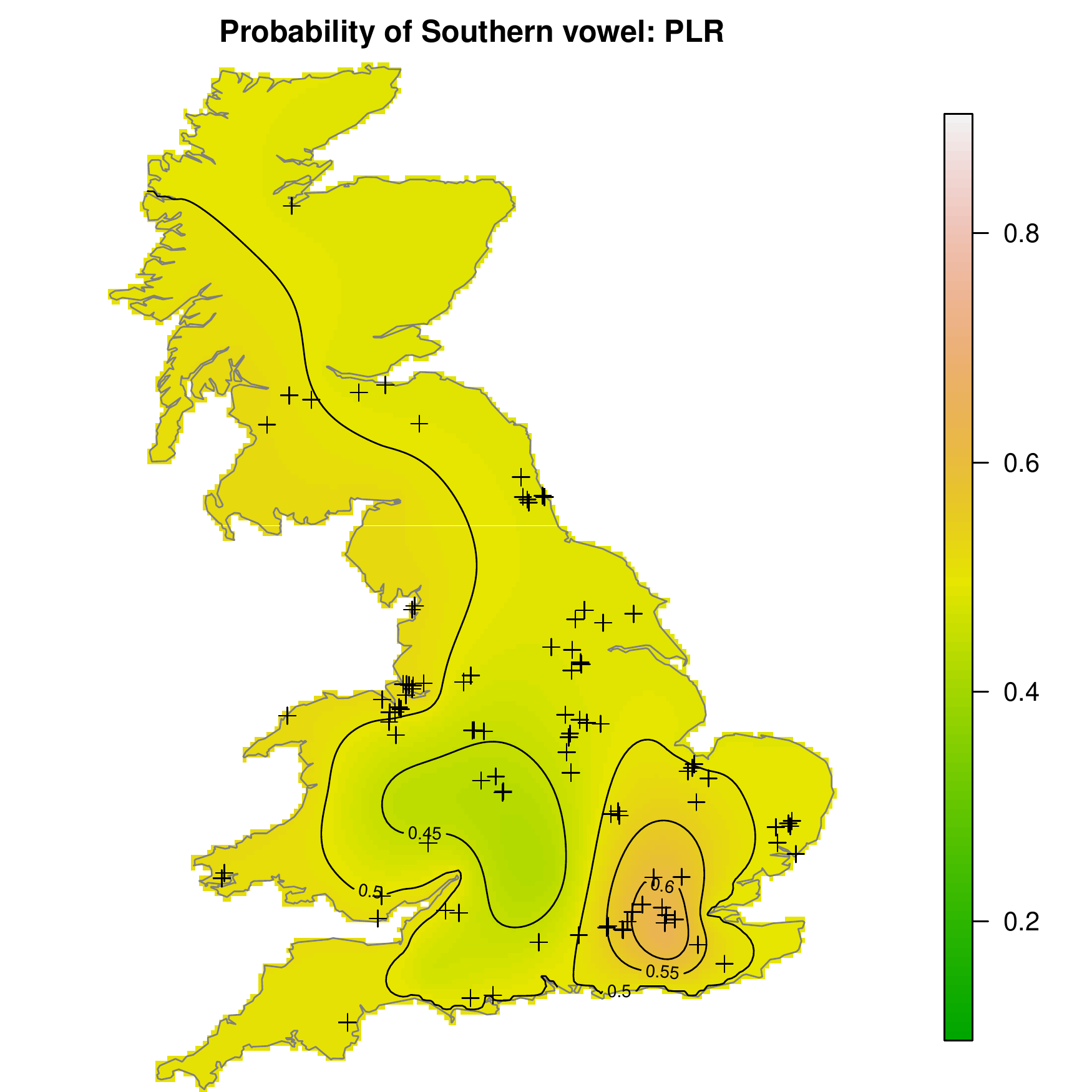}
        \caption{Map using MFCCs.}
    \end{subfigure}
    \caption{Smoothed predicted probabilities of a vowel sound being Southern, when using the two models of Section~\ref{sec:classify}. Black crosses are recording locations.}
    \label{fig:accent_maps}
\end{figure}

Plots of the fitted response surfaces $\hat \mu(\texttt{lon}, \texttt{lat}) = \ilogit( \hat f(\texttt{lon}, \texttt{lat}) )$ using the formant and the MFCC classification models are given in Figure~\ref{fig:accent_maps}. 
Both maps seem to suggest a North against Southeast split, similar to the isogloss map in Figure~\ref{fig:isogloss}. The predicted probabilities are usually not close to 0 or 1, because the BNC contains more variation than we have in the NSCV training data, due for instance to the variation in recording environments, and since not all speakers have a stereotypical Northern or Southern accent. 


\begin{figure}[p]
    \centering
    \begin{subfigure}{\textwidth}
        \centering
        \includegraphics[height=0.4\textheight]{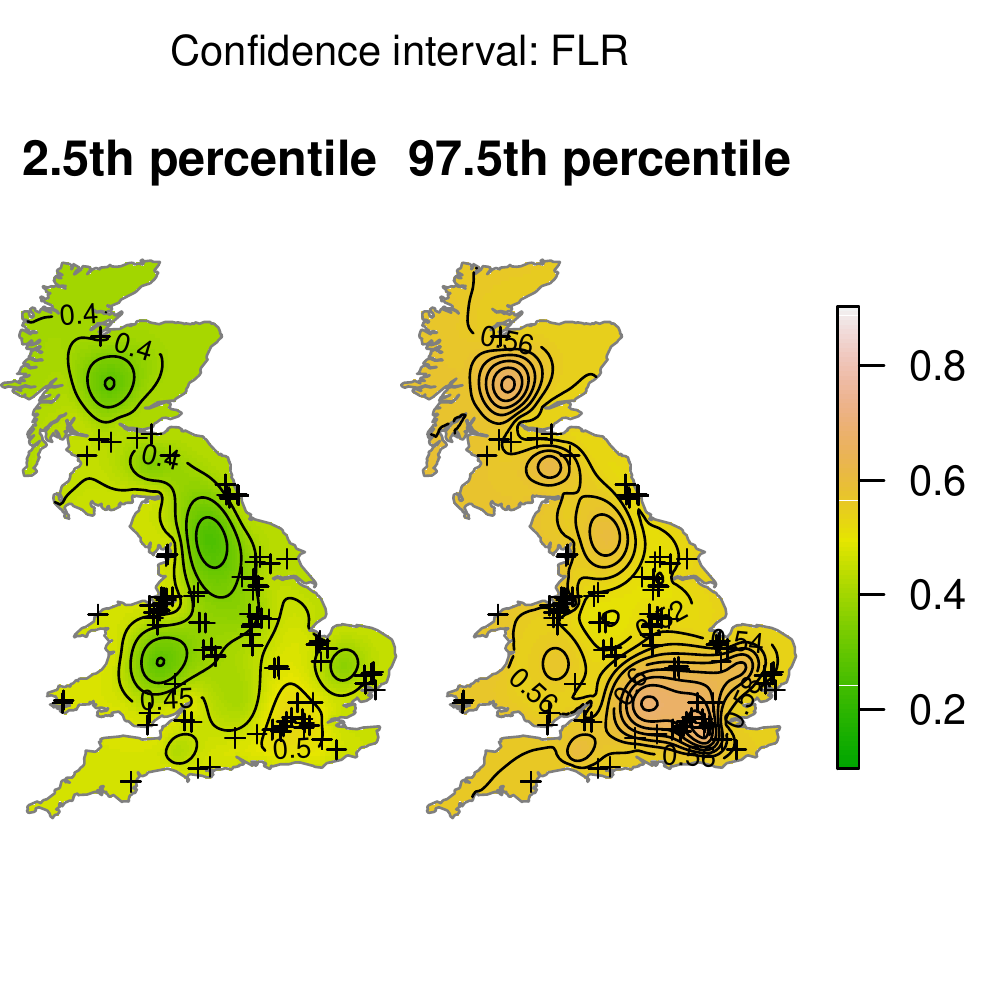}
        \caption{Pointwise confidence intervals for $\mu(\cdot, \cdot)$ for the formant model.}
    \end{subfigure}
    \begin{subfigure}{\textwidth}
        \centering
        \includegraphics[height=0.4\textheight]{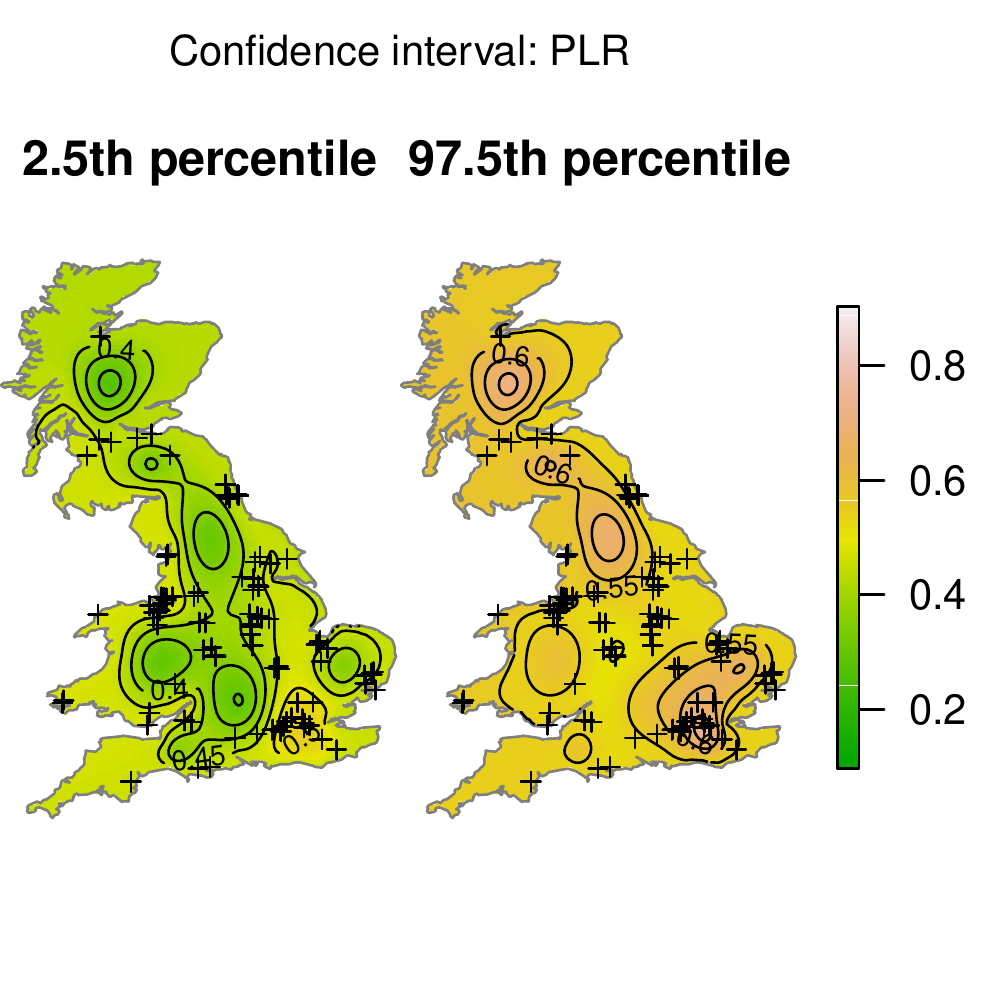}
        \caption{Pointwise confidence intervals for $\mu(\cdot, \cdot)$ for the MFCC model.}
    \end{subfigure}
    \caption{Contours of the spatially smoothed probabilities, showing the lower and upper bounds of a 95\% pointwise confidence interval for $\mu(\cdot, \cdot)$, constructed using a pointwise Normal approximation on the logit scale.}
    \label{fig:accent_SE_maps}
\end{figure}

To visualise the uncertainty associated with the contours in Figure~\ref{fig:accent_maps}, Figure~\ref{fig:accent_SE_maps} shows the approximate 95\% pointwise confidence intervals for $\mu$. These are computed as $[\ilogit( \hat{f} - 1.96\times \hat{\texttt{se}}(\hat{f})), \ilogit(\hat{f} + 1.96\times \hat{\texttt{se}}(\hat{f}))]$, based on a normal approximation on the link function scale. Notice that the uncertainty for both models is high in Northern England, Scotland and Wales, due to fewer observations in those regions. However, the North-Southeast variation is consistent and Greater London emerges as a region with significantly Southern accents.

\section{Discussion} \label{sec:discussion}

We first demonstrated two principled and interpretable approaches to modelling accent variation in speech sounds, using techniques from functional data analysis and generalised additive models. We presented a model that uses formant trajectories to classify \textit{bath} vowel sounds as Northern or Southern based on their similarity to /\ae/ and /\textipa{A/} vowels, trained on a set of labelled vowels\hl[]{ collected in an experimental setup}. The same audio dataset was also used in a different model using MFCC curves, by using functional principal components analysis to generate new features from the MFCC curves, and then classifying the sounds using $\ell_1$-penalised logistic regression of the FPC scores. We showed in Section~\ref{sec:resynthesising} how this MFCC model allowed us to resynthesise vowel sounds along a spectrum between /\ae/ and /\textipa{A}/.

These formant and MFCC models were used to predict the probability of a Southern accent for vowels from the audio BNC \citep{BNC}, our second dataset. The predictions were smoothed spatially to visualise the trap--bath split in England, Wales and Scotland, using a spatial beta regression with a soap film smoother. The resulting maps show a North versus South-east difference in accents which we can directly attribute to the variation in the /\ae/ or /\textipa{A}/ vowel quality of BNC sounds.

This analysis demonstrates how we can combine information from a labelled audio dataset such as the NSCV dataset, with the unlabelled BNC dataset. Despite the small sample of 4 speakers in the NSCV dataset, it allowed for vowel classification models to be trained. From cross-validation it seems that these classification models are highly accurate, a property that we believe would hold in similar recording conditions (such as background noise level) as the training data. 
However, the classifiers can only distinguish between Northern and Southern BNC \textit{bath} vowels to the extent that they differ by the /\ae/ and /\textipa{A}/ vowels captured in the NSCV training dataset. To produce a more valid characterisation of accent variation, one could use a labelled dataset of speech recordings from a larger corpus of speakers who can produce both accents accurately. Another limitation of this analysis is that we cannot verify the assumption of smooth spatial accent variation since we have no accent labels for BNC sounds. An extension of this work could involve augmenting the BNC by having human listeners manually classify a random sample of BNC vowels as Northern or Southern. These labels could then be used to train accent classifiers directly on BNC vowels, and also to validate the assumption of smooth spatial accent variation.



In phonetics, an ongoing research question has been whether dynamic information about formants is necessary for differentiating between vowels, or whether formant values at the midpoint of the vowel or sampled at fewer time points are sufficient \citep{watson1999, strange1983}. In Appendix \ref{app:modelselection} we have compared the functional formant model \ref{eq:loggam} to simpler models using $\text{F}_1$ and $\text{F}_2$ formants measured at the middle of the vowel, or at 25\%, 50\% and 75\% of the vowel. Even though the \textit{bath} vowel is a monophthong which doesn't contain significant vowel transitions, we see that the functional models show slight increases in cross-validated classification accuracy as compared to their non-functional versions, and sampling formants at more time points does not hinder classification. This is due to the regularisation of smooth terms in the generalised additive models used. The functional modelling approach also doesn't require specific time points for sampling to be chosen in advance, so it can be easily used with other vowels which have different formant trajectories. The MFCC model is another higher dimensional approach to modelling variation in vowels, as it uses information from more frequency bands. It has a slightly lower accuracy than the functional formant model, but its ability to resynthesise vowel sounds may be desirable for some applications.



We also observe some differences between the predictions of the formant model and the MFCC model. These models agree on vowel classification about 94\% of the time for the NSCV vowels and 73\% of the time for BNC vowels. The disagreements occur when the vowel quality is not clear, for example when formant curves are close to the boundary between the two vowels, or in the BNC, when there is considerable background noise and more variation in conversational speech. Nevertheless, the resulting spatial maps (Figure~\ref{fig:accent_maps} and Figure \ref{fig:accent_SE_maps}) show many similarities. Another way to compare the two models is to resynthesise vowels along a spectrum using the MFCC model, and classify these new vowels using the formant model. Changing the ``Southernness'' of a vowel with the MFCC model does change the corresponding prediction from the formant model, suggesting that similar information about vowel quality is being used by both models. We have added more detail on comparing the models in Appendix \ref{app:comparing-models}.

It is also possible to combine both MFCCs and formants from the same sound in one model. This can be done similarly to the MFCC model \ref{eq:PLR}, by appending the matrix of formant curves to the matrix of MFCC curves from each sound, and performing FPCA and $\ell_1$-penalised logistic regression as before. The disadvantage of this model is that we can neither interpret it from a speech production perspective (since it contains MFCCs which don't have a physical interpretation), nor use it to resynthesise vowels (since we cannot resynthesise vowels using formants). We have nevertheless trained this model, which has an accuracy of 92.75\%, and the results are in Appendix \ref{app:combined-model}.

The functional approach to modelling accent variation which we have demonstrated can easily be used with a larger corpus with more words or speakers. It can also be applied to to other vowels, including diphthongs (vowels which contain a transition, such as in ``house'') to visualise other accent variation in Great Britain, or other geographic regions. 

\section*{Supplementary Material}

\textbf{Data and R code:} The R code and preprocessed data used to generate these results can be obtained \hl[from the authors, on request]{online at https://doi.org/10.5281/zenodo.4003815}.

\noindent
\textbf{Other outputs:} \texttt{nscv.gif} shows animated formant trajectories of the NSCV data. Resynthesised vowels are in \texttt{class-NtoS.wav} (perturbing the vowel in ``class'' from /\ae/ towards /\textipa{A/}), and \texttt{blast-StoN.wav} (perturbing the vowel in ``blast'' from /\textipa{A/} towards /\ae/).

\section*{Acknowledgements}

We thank the Associate Editor and three referees for their comments that helped to improve the quality of the paper.


\bibliographystyle{agsm}
\citestyle{agsm}
\bibliography{references-final.bib}


\pagebreak
\appendix

\section*{Appendices}
\section{Exploration of BNC}
\label{app:exploration}

\begin{figure}[p]
    \centering
    \includegraphics[width=\textwidth]{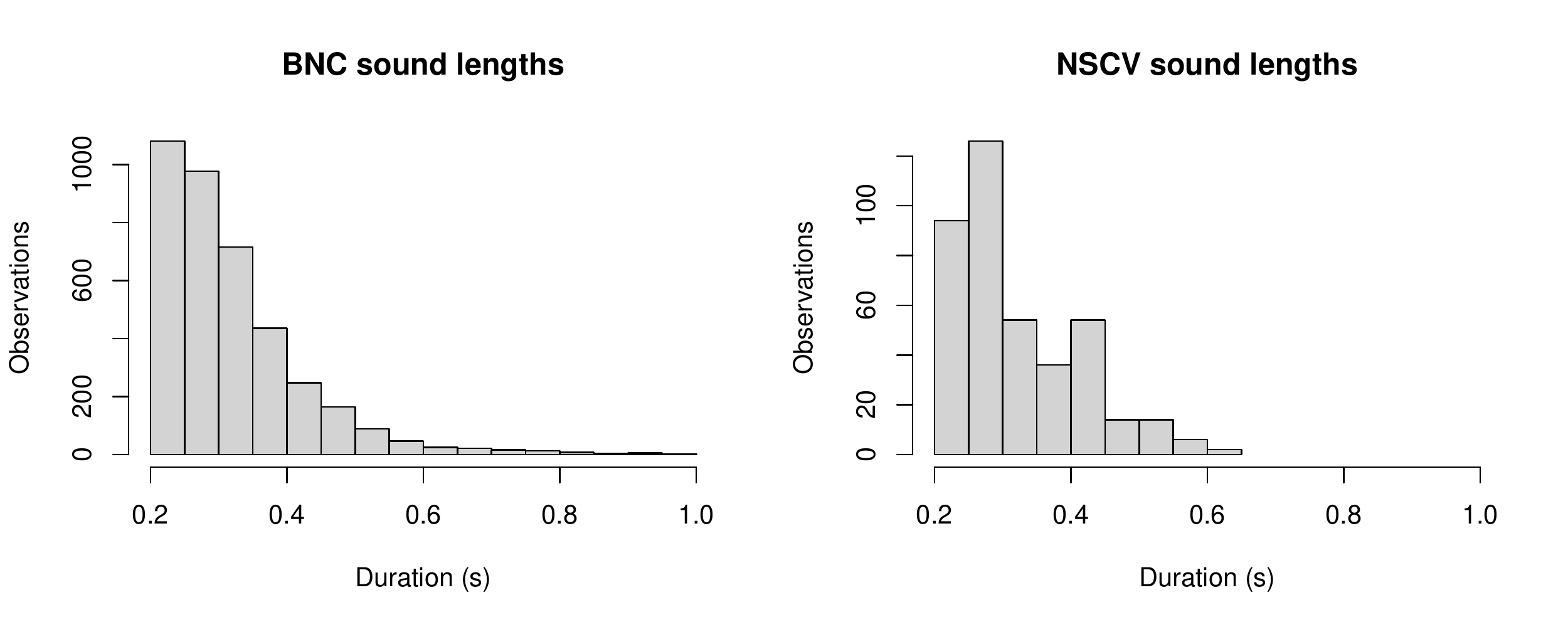}
    \caption{Histogram of sound lengths in the NSCV and BNC datasets.}
    \label{fig:bncdurations}
\end{figure}

\begin{figure}[p]
    \centering
    \includegraphics{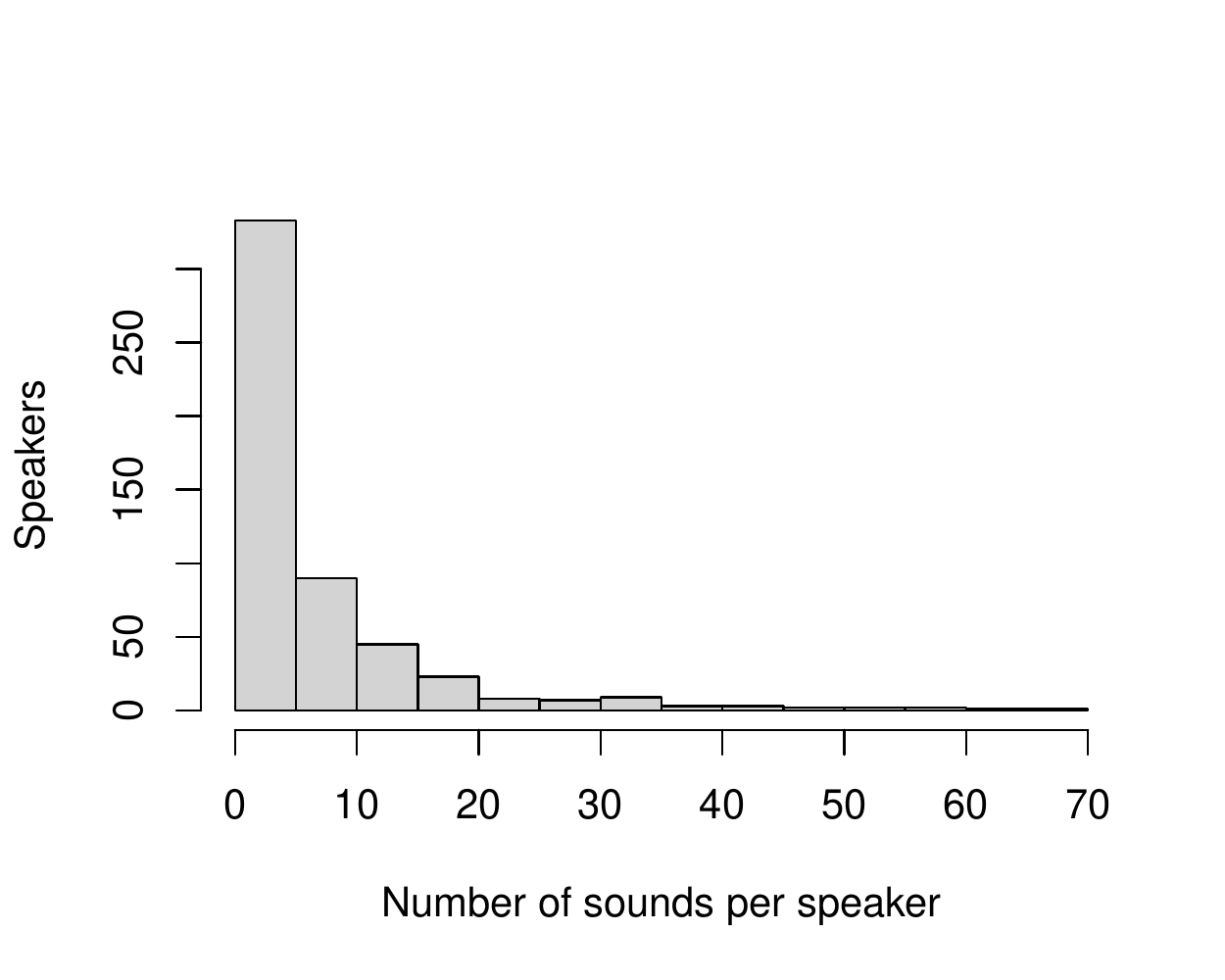}
    \caption{Histogram of number of utterances from each speaker in the set of ``class'' words analysed from the BNC.}
    \label{fig:sounds_per_speaker}
\end{figure}

Each observation in the dataset is a recording of a single utterance of a ``class'' word. The metadata that we have for each observation is shown in Table~\ref{table:bnc}, along with the criteria used to clean the data. Since our interest is in the differences in natural native accents, we clean the dataset using the available metadata, to remove observations from trained speakers such as newsreaders, and speakers with foreign accents. We also remove observations from children, since the acoustic properties of their speech is considerably different from adult speech \citep{Safavi2018}. We limit our analysis to accents in Great Britain, and therefore removed recordings from Northern Ireland. Words that are shorter than $0.2$ seconds or longer than $1$ second were removed. 
\hl[]{We also removed the most noisy sounds using the procedure described in the online supplement of \citet{Tavakoli2019}.}
Histograms of the sound lengths in the BNC and NSCV are in Figure~\ref{fig:bncdurations}. 
The cleaned dataset contains 3852 clips from 529 speakers in 124 locations across England, Scotland and Wales. Figure~\ref{fig:obsnum} shows the number of observations per locations, Figure~\ref{fig:sounds_per_speaker} shows a histogram of utterances per speaker, and Table~\ref{table:bnc_social} gives a breakdown of the social classes recorded in the metadata for the list of ``class'' words analysed.

\begin{table}[hb]
    \scriptsize
    \centering
    \begin{tabular}{@{} l  p{6cm}  p{6cm} @{}} 
        \toprule
        Covariate & Values & Cleaning criteria \\ [0.5ex] 
        \midrule
        Sex & 39\% female, 43\% male and 18\% missing &  None \\ 
        Age & Ages 2-95 years & Removed speakers below 10 years old. \\
        Social class & Four categories designated according to occupation, with 57\% unknown &  None \\
        Recording location & Location of recording eg. ``Bromley, London'' & Removed Northern Ireland, unknown locations, and speakers whose dialect was not native to their recording location.\\
        Occupation & Name of the profession, 23\% unknown & Removed trained professional speakers, eg. football commentators, radio presenters, newsreaders.\\ 
        Activity & Activity during recording, 92\% unknown & Removed activities like radio and TV broadcasts. \\
        Duration of word & Between 0.09 and 5.4 seconds & Kept words between 0.2 and 1 second long. \\
        Dialect & Regional dialect categories, 50\% unlabeled & Removed speakers with known foreign accents eg.\ Indian, American and Chinese. \\
        \bottomrule
    \end{tabular}

    \caption{Covariates in the BNC and the criteria used for including observations in our analysis.}
    \label{table:bnc}
\end{table}

\begin{table}[ht]
    \centering
        \begin{tabular}{@{} lll @{}} 
            \toprule
            Social Class & Code & Number of speakers \\ [0.5ex] 
        \midrule
            Upper middle class & AB  & 48 \\ 
            Lower middle class & C1  & 64 \\
            Skilled working class & C2  & 54 \\
            Working class and non working & DE  & 46 \\
            Unknown & UU & 317 \\ 
        \bottomrule
    \end{tabular}

    \caption{Number of speakers recorded in each social class in the BNC metadata for the list of ``class'' words analysed.}
    \label{table:bnc_social}
\end{table}

\section{Model selection for formant model}
\label{app:modelselection}

Here are some descriptive statistics and plots of the NSCV formants, and the results from the model selection procedure. 

\subsection{Formants} \label{app:formant-eda}
\begin{figure}[h]
    \centering
    \includegraphics[width=\textwidth]{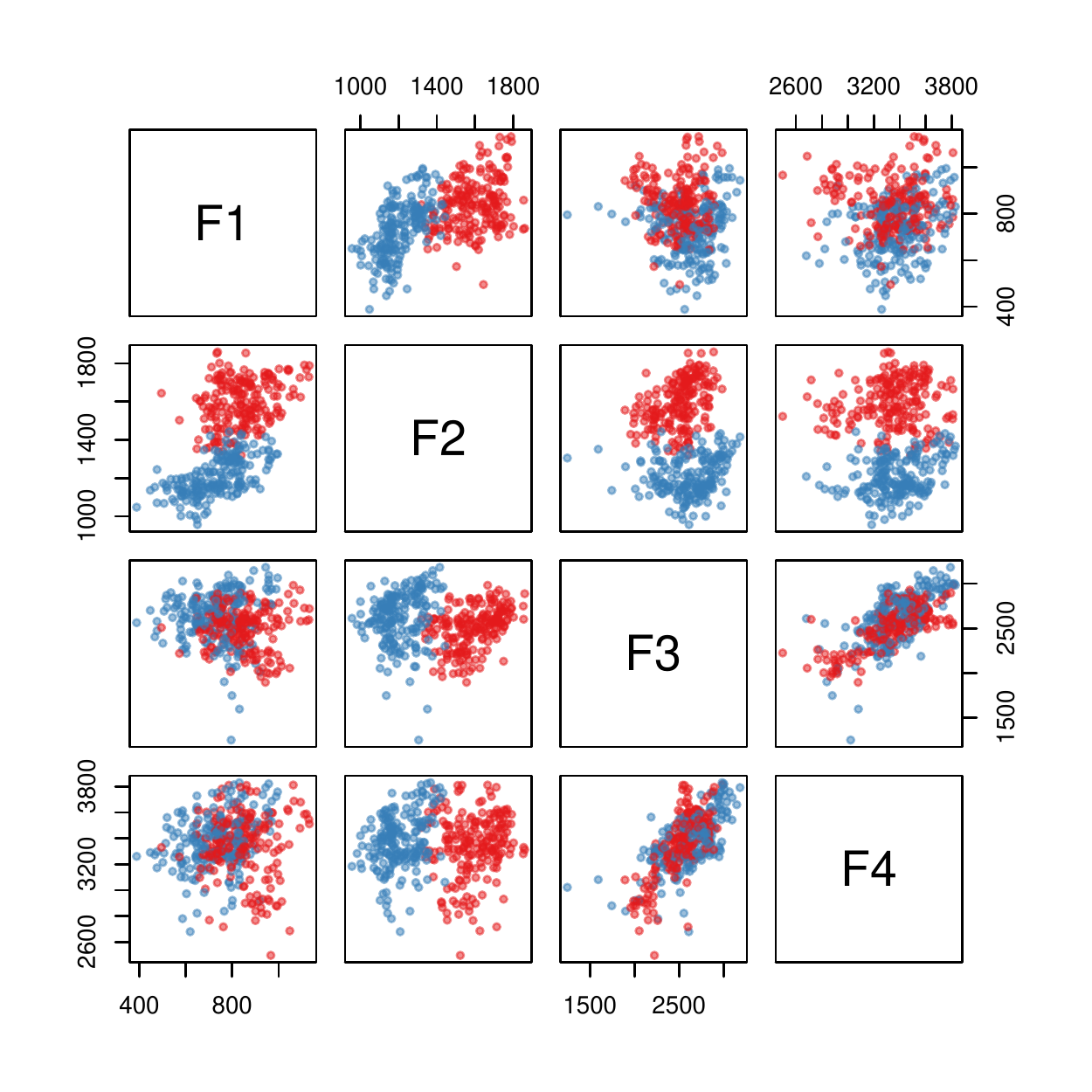}
    \caption{Plot of average formant values from each sound in the NSCV dataset. Northern accents are shown in blue and Southern in red.}
    \label{fig:eda_formant_avgs}
\end{figure}

\begin{figure}[h]
    \centering
    \includegraphics[width=\textwidth]{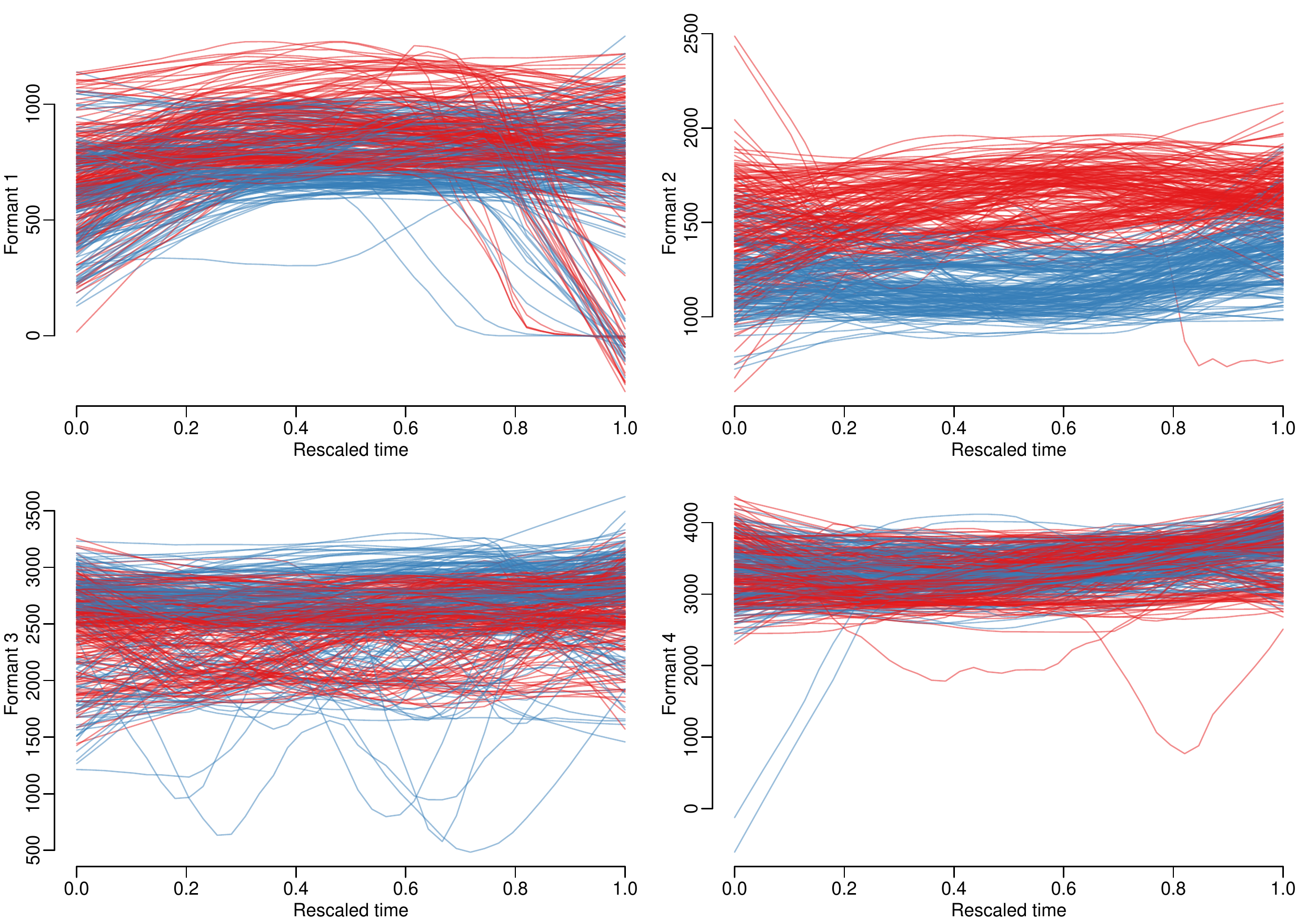}
    \caption{Plot of formant curves from each sound in the NSCV dataset. Northern accents are shown in blue and Southern in red.}
    \label{fig:eda_formant_curves}
\end{figure}

Table \ref{table:eda_formant_avgs} shows the average formant values for both vowels in the dataset, and Figure \ref{fig:eda_formant_avgs} plots each sound's average formant values against each other. Figure \ref{fig:eda_formant_curves} shows the entire smoothed formant curves.

\begin{table}[hb]
    \caption{Average formant values in the NSCV dataset}
    \label{table:eda_formant_avgs}
    \centering
    \begin{tabular}{rcrrrr}
        \toprule
        \textbf{Accent} && $\text{F}_1$ & $\text{F}_2$ & $\text{F}_3$ & $\text{F}_4$ \\
        \cmidrule{1-6} Northern && 841 & 1595 & 2504 & 3344 \\
        Southern  && 731 & 1206 & 2627 & 3376 \\
        \bottomrule
    \end{tabular}
\end{table}

\subsection{Model selection} \label{app:formant-model-selection}

Here are the results from the model selection procedure used to select the terms in the formant model. Table \ref{table:model_selection2} shows models that were trained on NSCV formants after aligning all NSCV MFCC 1 curves together, as described in Section \ref{sec:preprocessing}. All models also had random effects for each speaker. We compare the effective degrees of freedom, AIC and cross validated accuracy for each model.

The first three models use single formant values taken from the middle of the sound, models 4--6 use formant values taken at 3 points along the vowel (at 25\%, 50\% and 75\% of the sound), and models 7--11 use the entire curve containing 40 samples from each curve. The models with formants at 3 time points are fitted in the same way as model \ref{eq:loggam}, with the 3 formant values from each curve contributing to the predictor through a linear functional term, and the coefficient curves represented with a cubic spline with penalised second derivative.  We can see that the functional models usually have slightly higher accuracy and better fit to the data than their non-functional equivalents, and that increasing the number of samples from 3 to 40 doesn't meaningfully improve the classification accuracy.

In the paper we use the functional model 8 in Table \ref{table:model_selection2} to classify vowels based on the $\text{F}_2$ curve, as it has the lowest AIC, second highest accuracy, and relatively low degrees of freedom.

\begin{table}[hb]
    \caption{Formant model selection results, using aligned formant curves. All models include speaker random effects as in model \ref{eq:loggam} in Section \ref{sec:formant-model}. We compare effective degrees of freedom (EDF), adjusted AIC and cross-validated accuracy.}
    \label{table:model_selection2}
    \centering
    \begin{tabular}{lrrrrr}
        \toprule
        \textbf{Model no.} & Fixed Effects && EDF & AIC & Accuracy \\
        \cmidrule{1-6}
        1 & $\text{F}_1$ at middle of vowel && 4.98 & 227.50 & 66.75\% \\
        
        2 & $\text{F}_2$ at middle of vowel && 4.24 & 53.62 & 96.5\% \\
        
        3 & $\text{F}_1$ + $\text{F}_2$ at middle of vowel&& 5.69 & 48.17 & 89.75\% \\
 
        4 & $\text{F}_2$(t) at 3 time points && 4.13 & 26.88 & 97\% \\
        
        5 & $\text{F}_1$(t) + $\text{F}_2$(t) at 3 time points && 5.00 & 22.61 & 98.75\% \\
        
        6 & mean($\text{F}_1$) + ($\text{F}_2$ - $\text{F}_1$)(t) at 3 time points && 6.30 & 58.22 & 95 \% \\
        
        7 & $\text{F}_1$(t) && 9.84 & 214.44 & 66.5\% \\
        
        8 & $\text{F}_2$(t) && 6.76 & 13.74 & 98\% \\
        
        9 & ($\text{F}_2$ - $\text{F}_1$)(t) && 10.51 & 122.16 & 88.5\% \\
        
        10 & mean($\text{F}_1$) + ($\text{F}_2$ - $\text{F}_1$)(t) && 9.19 & 18.38 & 96.25\% \\
       
        11 & $\text{F}_1$(t) + $\text{F}_2$(t) && 7.25 & 14.55 & 98\% \\

        \bottomrule
    \end{tabular}
\end{table}

In Table \ref{table:model_selection2_noalignment} we also show the same models trained on NSCV vowels before the alignment preprocessing step. For most of the models, the alignment has led to an increase in accuracy.
In our procedure the alignment of vowels in the NSCV corpus is an important step as this allows us to align the BNC vowels to the NSCV vowels before predicting accents for the BNC. In general if misalignments between the training dataset and prediction dataset are not a concern, one can still obtain a good classifier using formants without the alignment step.

\begin{table}[hb]
    \caption{Formant model selection results, without alignment of formant curves. As before, all models include speaker random effects and we compare effective degrees of freedom (EDF), adjusted AIC and cross-validated accuracy.}
    \label{table:model_selection2_noalignment}
    \centering
    \begin{tabular}{crrrrr}
        \toprule
        \textbf{Model no.} & Fixed Effects && EDF & AIC & Accuracy \\
        \cmidrule{1-6}
        1 & $\text{F}_1$ at middle of vowel && 4.98 & 272.76 & 65.25\% \\
        
        2 & $\text{F}_2$ at middle of vowel && 4.29 & 62.28 & 95.25\% \\
        
        3 & $\text{F}_1$ + $\text{F}_2$ at middle of vowel && 5.90 & 44.54 & 84\% \\
 
        4 & $\text{F}_2$(t) at 3 time points && 4.78 & 30.76 & 96.75\% \\
        
        5 & $\text{F}_1$(t) + $\text{F}_2$(t) at 3 time points && 8.72 & 17.51 & 94.5\% \\
        
        6 & mean($\text{F}_1$) + ($\text{F}_2$ - $\text{F}_1$)(t) at 3 time points && 6.71 & 50.46 & 95.25\% \\

        7 & $\text{F}_1$(t) && 8.59 & 258.57 & 61.5\% \\
        
        8 & $\text{F}_2$(t) && 7.06 & 14.36 & 96.25\% \\
        
        9 & ($\text{F}_2$ - $\text{F}_1$)(t) && 17.2 & 94.78 & 91.5\% \\
        
        10 & mean($\text{F}_1$) + ($\text{F}_2$ - $\text{F}_1$)(t) && 6.5 & 29.6 & 88\% \\
    
        11 & $\text{F}_1$(t) + $\text{F}_2$(t) && 6.83 & 13.72 & 94.5\% \\

        \bottomrule
    \end{tabular}
\end{table}

\section{A model combining MFCCs and formants}
\label{app:combined-model}

Here is a model similar to model \ref{eq:PLR}, in which we use both formant and MFCC curves together to classify vowels. In general applications usually use one form of speech representation at a time, as they would contain very similar information about the same sound. Following the same approach as the MFCC model, we first perform functional principal components analysis to reparametrise the 44 curves from each sound (40 MFCC curves and 4 formant curves), and use the resulting FPC scores in an $\ell_1$-penalised logistic regression model.

\begin{figure}[p]
    \centering
        \begin{subfigure}{\textwidth}
        \centering
        \includegraphics[height=0.4\textheight]{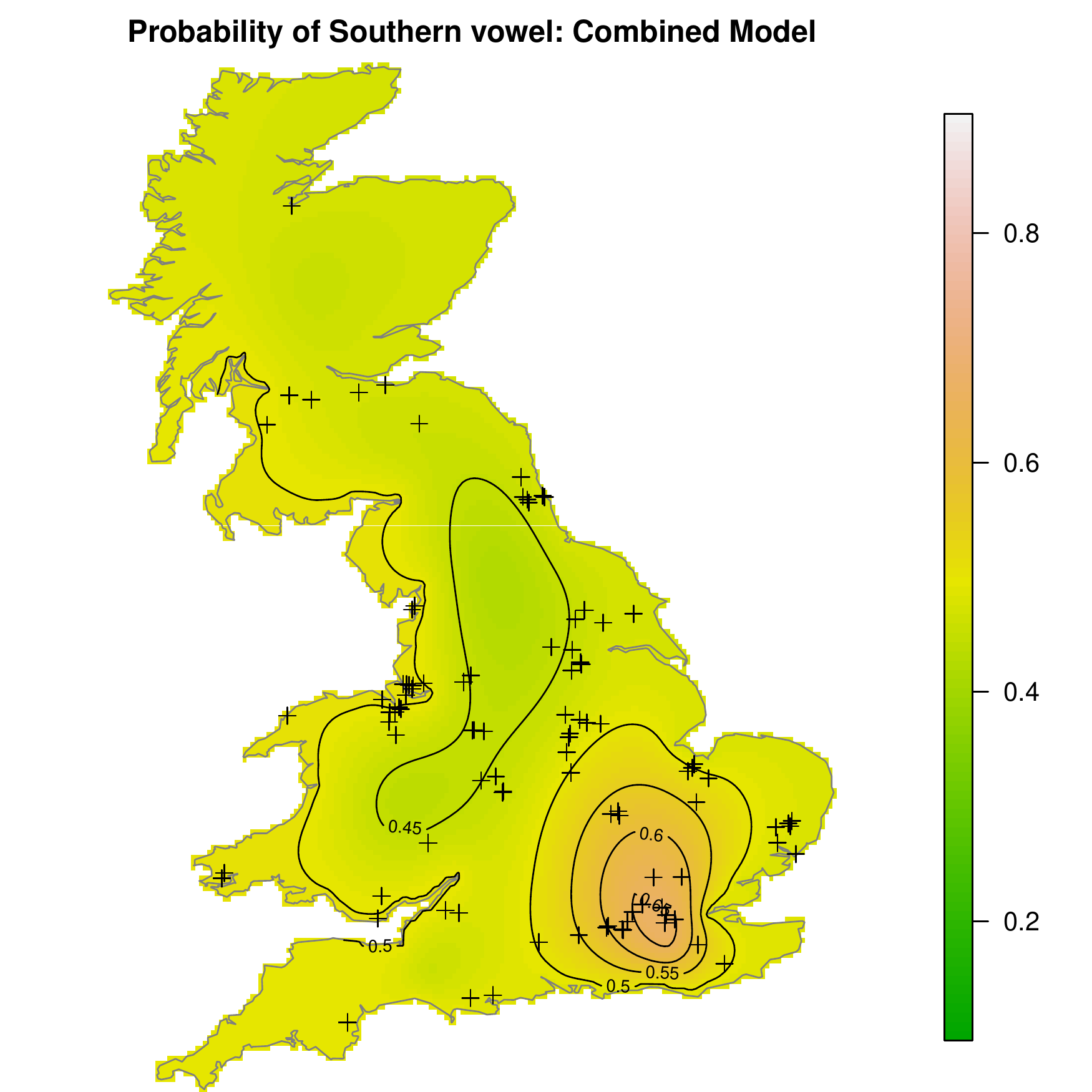}
        \caption{Spatially smoothed probabilities from the combined model with MFCCs and formants.}
    \end{subfigure}
    \begin{subfigure}{\textwidth}
        \centering
        \includegraphics[height=0.4\textheight]{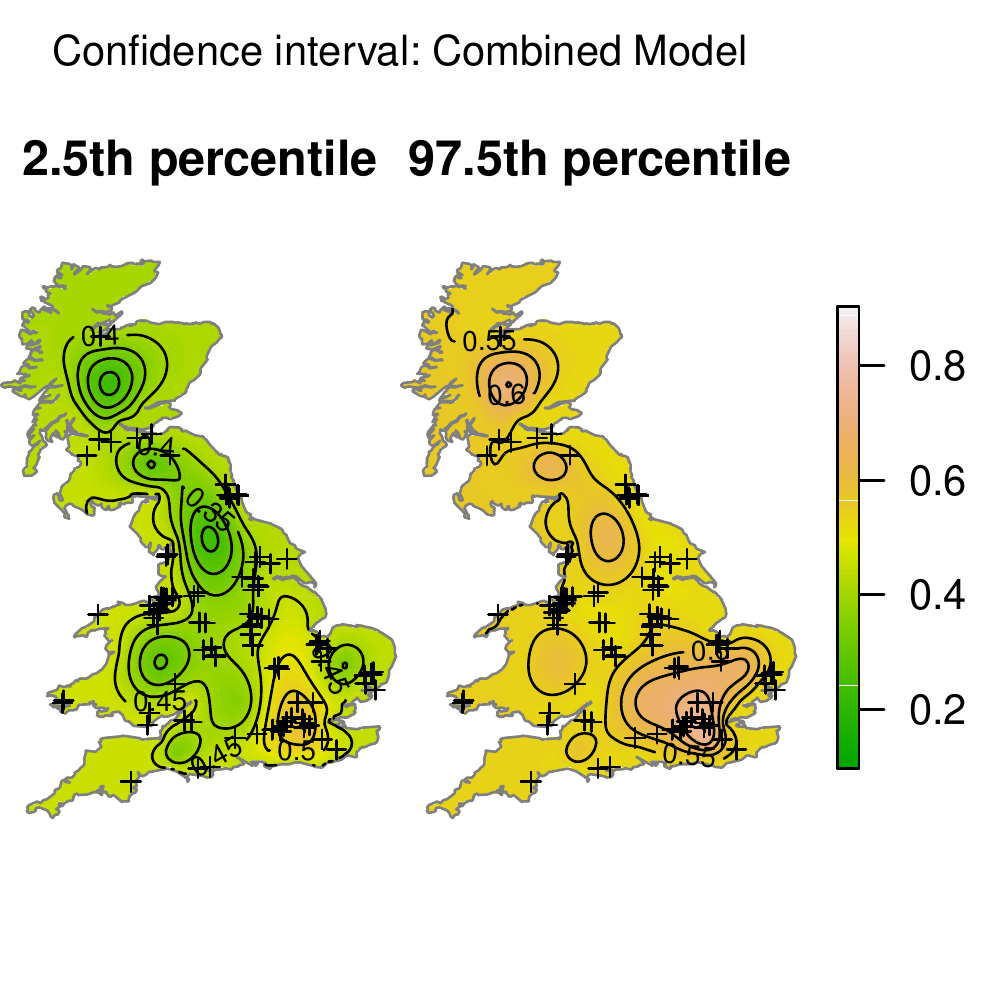}
        \caption{Pointwise confidence intervals from the combined model with MFCCs and formants.}
    \end{subfigure}
    \caption{Contours of the spatially smoothed probabilities, showing the lower and upper bounds of a 95\% pointwise confidence interval for $\mu(\cdot, \cdot)$, constructed using a pointwise Normal approximation on the logit scale.}
    \label{fig:combined-map}
\end{figure}

The cross validated accuracy of the model is 92.75\% and 32 FPCs have nonzero coefficients. The confusion matrix is shown in Table \ref{table:combined-conf} and the spatial map of smoothed BNC predictions and the confidence intervals maps are shown in Figure \ref{fig:combined-map}. The overall spatial variation is very similar to what we see in maps from the separate formant and MFCC models.

\begin{table}[h]
    \caption{Cross-validated confusion matrix for the classifier using both MFCCs and formants.}
    \label{table:combined-conf}
    \centering
    \begin{tabular}{rcrr}
        \toprule
        {} &\phantom{} &\multicolumn{2}{c}{\textbf{Truth}}\\
        \cmidrule{3-4} && North & South \\
        \textbf{Prediction} \\
        North && 191 & 20  \\
        South && 9 & 180 \\
        \bottomrule
    \end{tabular}
\end{table}

\section{Comparing the formant-based model and MFCC-based model}
\label{app:comparing-models}

The formant model \ref{eq:loggam} and MFCC model \ref{eq:PLR} sometimes classify the same vowel differently, which leads to some differences between the two spatial maps. In this Appendix we examine where these disagreements between the models occur.

\begin{figure}[p]
    \centering
    \includegraphics[width=\textwidth]{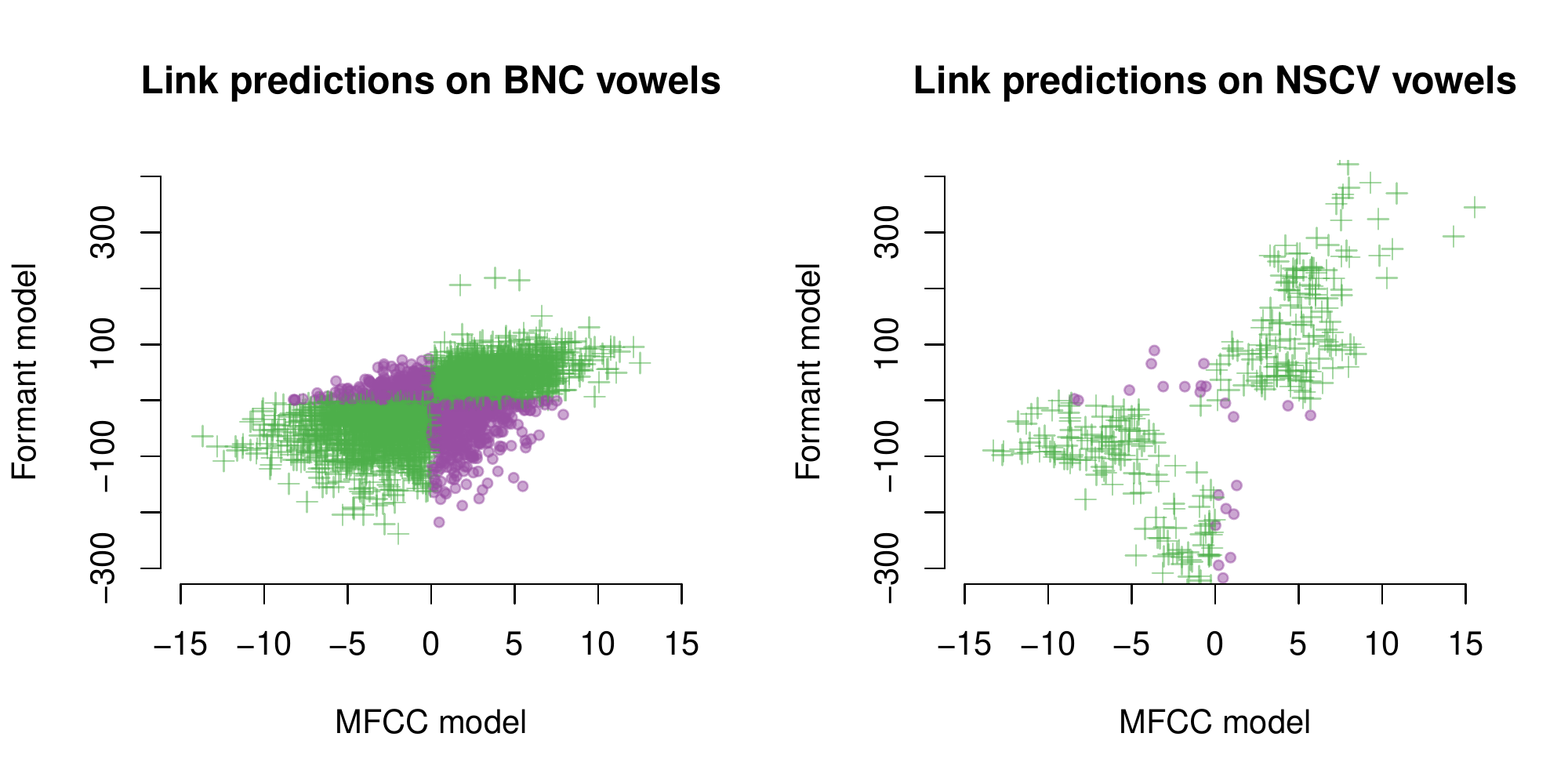}
    \caption{Comparing predictions on the link scale, from formant and MFCC models. Green crosses are observations where the models agree and purple dots are where they disagree. }
    \label{fig:comparison-link}
\end{figure}

Figure \ref{fig:comparison-link} shows the linear predictor values from the two models, when predicting the accent of NSCV vowels and BNC vowels. The models agree on the classification 94\% of the time for NSCV vowels (when the vowels being predicted are not part of the training set), and 73\% of the time for BNC vowels. The BNC predictions from both models on the linear predictor scale are closer to 0 than the NSCV predictions, because the BNC vowels are not as clearly differentiated as the vowels collected under controlled experimental conditions.

\begin{figure}[h]
    \centering
    \includegraphics[width=\textwidth]{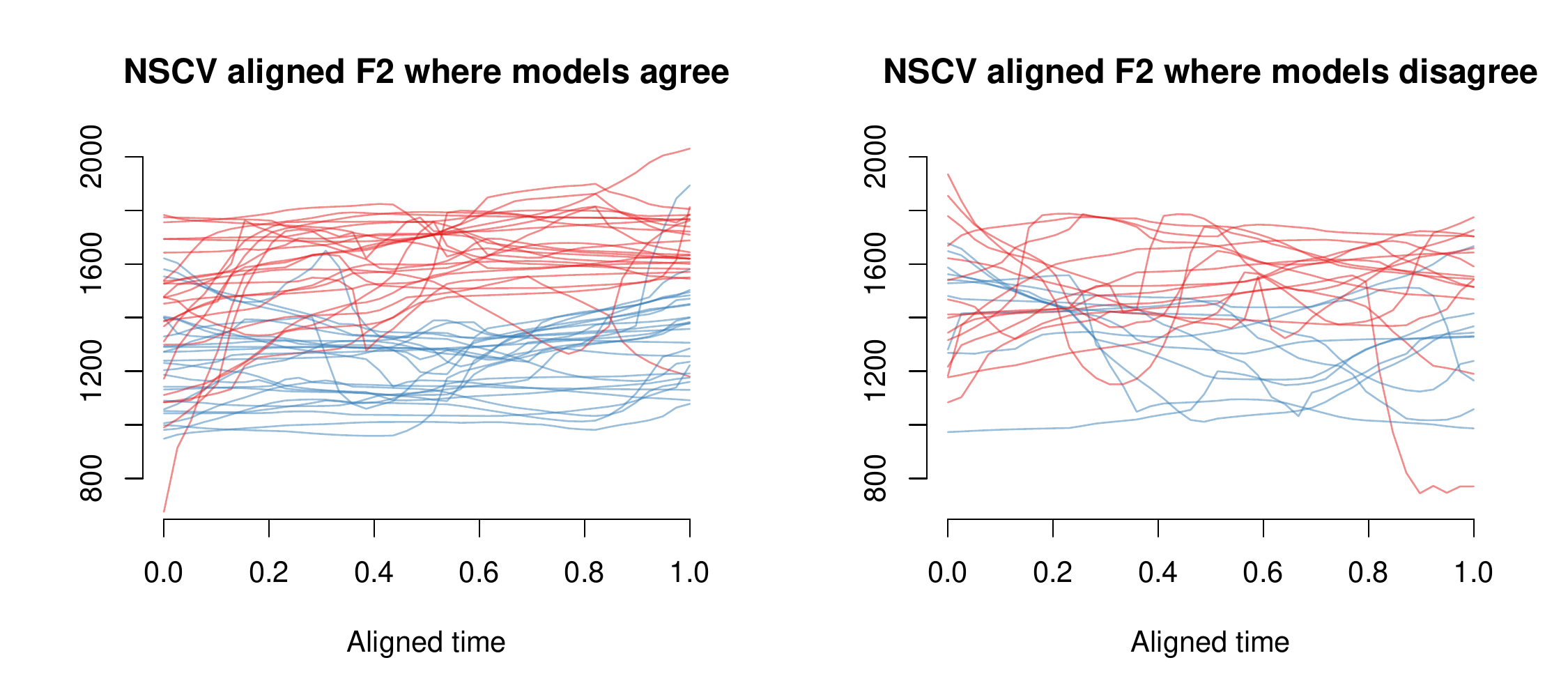}
    \caption{Comparison of NSCV $\text{F}_2$ curves where the models agree and disagree. Red curves are Southern and blue curves are Northern.}
    \label{fig:comparison-nscv}
\end{figure}

\begin{figure}[h]
    \centering
    \includegraphics[width=\textwidth]{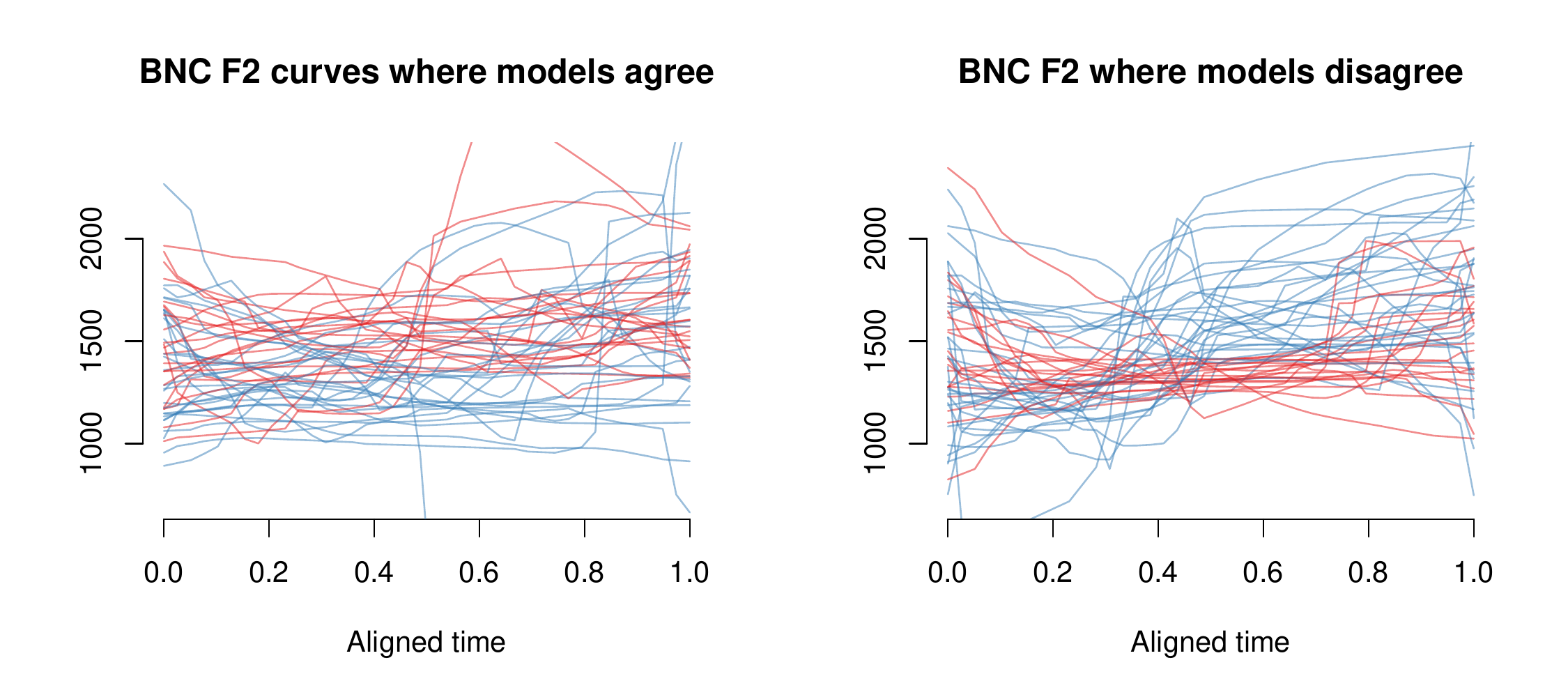}
    \caption{Comparison of BNC $\text{F}_2$ curves where the models agree and disagree. Red curves are predicted Southern by the MFCC model and blue curves are predicted Northern.}
    \label{fig:comparison-bnc}
\end{figure}

We can also look at some features of the vowels themselves. Figure \ref{fig:comparison-nscv} shows a sample of 50 $\text{F}_2$ curves from NSCV vowels where the models agree and all 25 curves where they disagree. The observations where the models disagree are not as clearly separated. A similar plot for BNC vowels is in Figure \ref{fig:comparison-bnc}, coloured by their classifications under the MFCC model (since we don't know the ground truth).

A third way of comparing the models is to resynthesise new vowels using the MFCC model, and classify the new vowels using the formant model. This allows us to compare one model's predicted probabilities against the other. 

\begin{figure}[h]
    \centering
    \includegraphics[width=6in]{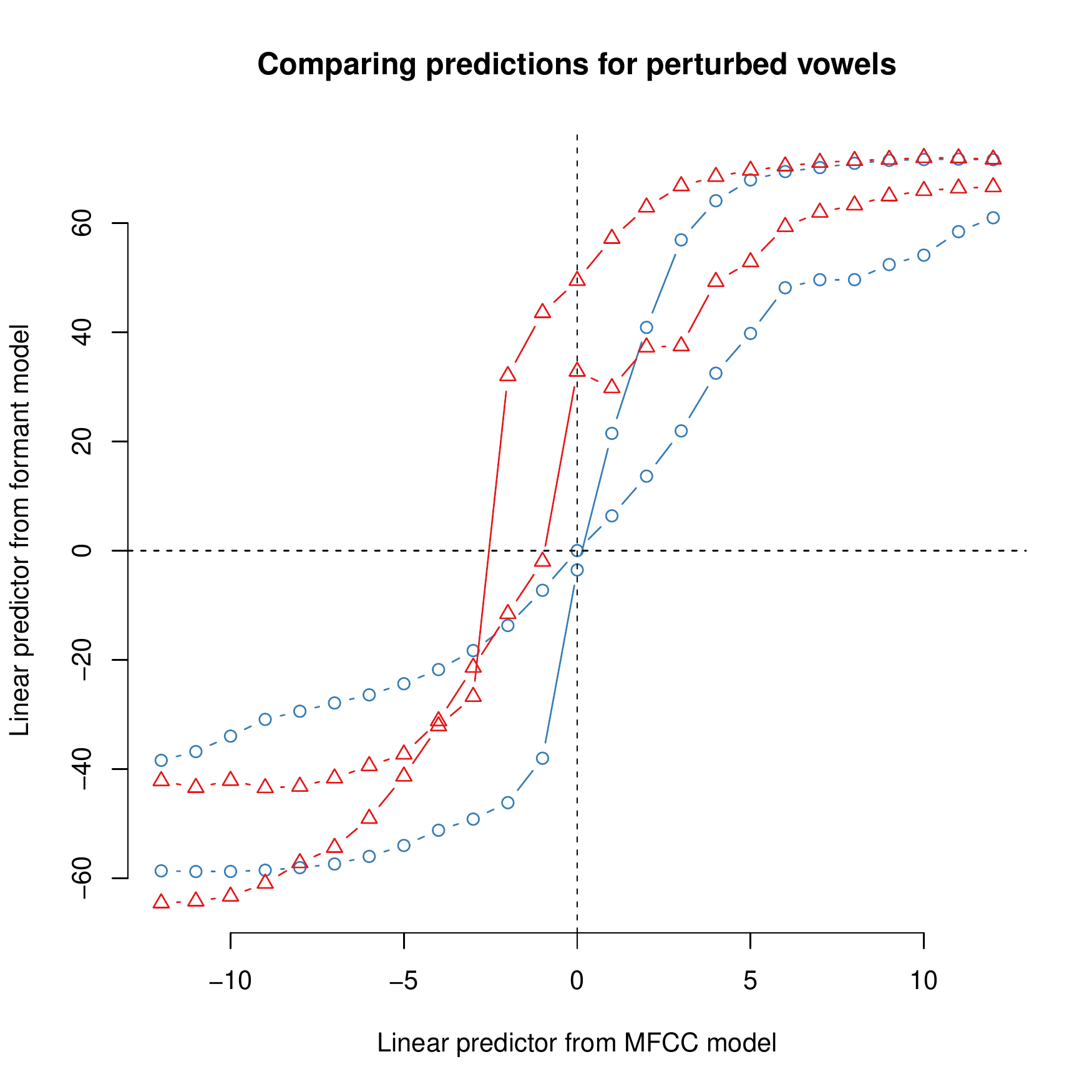}
    \caption{Predictions on the link scale by the formant model, for 4 vowels resynthesised with the MFCC model. The blue lines with circles correspond to vowels that were originally Northern, that were resynthesised along a grid towards a Southern accent. The red lines with triangles corresponds to vowels that were originally Southern and were resynthesised towards a Northern accent.}
    \label{fig:comparison-perturbed-link}
\end{figure}

\begin{figure}[h]
    \centering
    \includegraphics[width=6in]{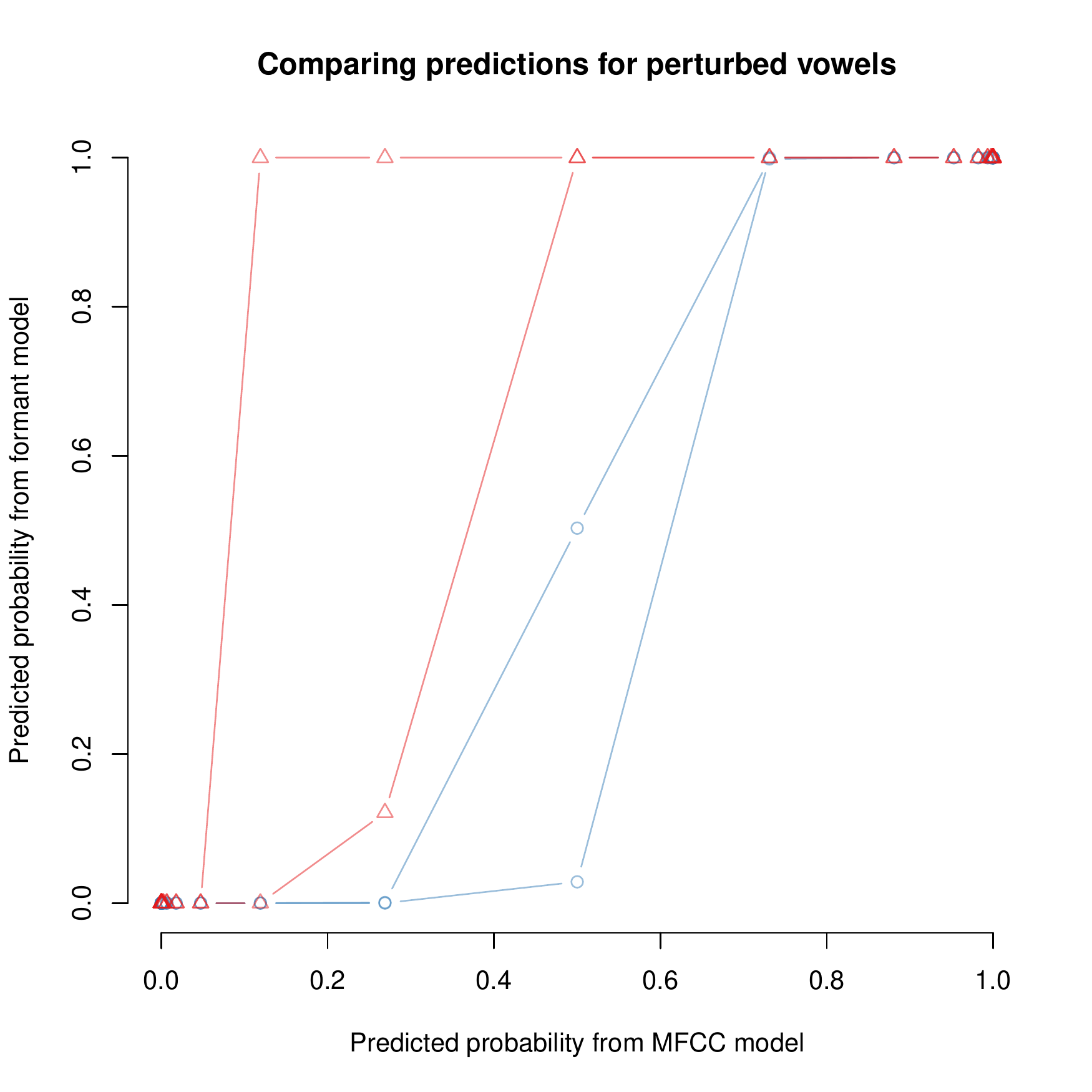}
    \caption{Predicted probabilities under the formant model, of vowels resynthesised by the MFCC model, as in Figure \ref{fig:comparison-perturbed-link}.}
    \label{fig:comparison-perturbed-prob}
\end{figure}

This is done similarly to the procedure in Section \ref{sec:resynthesising}. We begin with a chosen vowel sound, preprocess its MFCCs, and predict its probability of being Southern under the MFCC model. Using this prediction on the linear predictor scale, we can calculate the sequence of multiples of the MFCC matrix (Figure \ref{fig:contrib}) to add to the sound in order to achieve certain predictions under the MFCC model. We can add these MFCCs to the original MFCCs and synthesising the resulting sound to obtain a new sequence of vowels. We then extract and preprocess the formants from these sounds, and calculate their predictions under the formant model. Examples of this for two Northern and two Southern vowels are shown in Figure \ref{fig:comparison-perturbed-link} where we perturb vowels between $-12$ and $12$ on the MFCC linear predictor scale. Figure \ref{fig:comparison-perturbed-prob} shows the same plot on the probability scale.

This shows us that the formant model validates the vowel change that we can create using the MFCC model. It suggests that the sound information being used by both these models is closely related.

\end{document}